\documentclass[12pt]{article}
\usepackage{a4,epsfig}
\begin{document}




\begin{titlepage}

\pagenumbering{arabic}
\vspace*{-1.5cm}
\begin{tabular*}{15.cm}{lc@{\extracolsep{\fill}}r}
&
\\
& &
30 October 2017  
\\
&&\\ \hline
\end{tabular*}
\vspace*{2.cm}
\begin{center}
\Large 
{\bf \boldmath
Looking for a Theory              \\
of Faster-Than-Light Particles     
} \\
\vspace*{2.cm}
\normalsize { 
   
   {\bf V.F. Perepelitsa}\\
   {\footnotesize ITEP, Moscow            }\\ 
}
\end{center}
\vspace{\fill}
\begin{abstract}
\noindent
Several principal aspects of a theoretical approach to the theory of
faster-than-light particles (tachyons) are considered in this note.
They concern the resolution of such problems of tachyon theory as
the causality violation by tachyons, the stability of the tachyon vacuum,
and the stability of ordinary particles against the spontaneous emission of 
tachyons, i.e. the problems which are generally used as arguments against
the possibility of such particles. It is demonstrated that all these arguments 
contain nontrivial loopholes which undermine their validity.
A demand for a consistent tachyon theory is formulated, and 
several ideas for its construction are suggested. 

\end{abstract}
\vspace{\fill}
                    
\vspace{\fill}
\end{titlepage}




\setcounter{page}{1}    


\section{Introduction}
This note presents a consideration of several problems related to
faster-than-light particles: particles with spacelike momenta, called tachyons.
They are listed below:
\begin{itemize}
\item [1.] Violation of causality by tachyons.
\item [2.] Instability of the tachyon vacuum.
\item [3.] Violation of unitarity by interacting tachyons.
\end{itemize}
Usually these problems serve as arguments against the possibility of the
existence of such particles. Each of the arguments considered separately 
looks valid, but taken together they, most probably, are wrong.
   
It is the aim of this note to show how all these problems can be solved in the
frame of standard physics via a non-standard approach. This approach, 
formulated in parallel with the problems under consideration, 
is based on the properties of faster-than-light particles which look 
quite natural from an unbiased point of view, though separate tachyons into 
a class of rather unusual objects in the particle world.  

The first theoretical arguments for the possibility of the existence of 
particles with spacelike momenta\footnote{We denote 4-vectors by ordinary 
italic type; boldface type is used to denote 3-vectors when it is necessary to
avoid confusion.}, 
$P^2 < 0$, can be found in a famous paper by Wigner in which the classification 
of unitary irreducible representations (UIR's)
of the Poincar\'{e} group was done for the first time~\cite{wigner1}.
In this work Wigner used for the classification of the UIR's two Casimir
invariants of the Poincar\'{e} group, one of which is the particle 
four-momentum squared, $P^2$. Three classes of the UIR's were distinguished
by the values of this Casimir invariant, one of them having $P^2 > 0$
corresponding to ordinary massive particles, the second class having $P^2 = 0$
corresponding to massless particles (to photons and probably to some sort of
neutrinos\footnote{Indeed this class was presented in \cite{wigner1} by two
subclasses: $P^2 = 0$ with not all P components being zero, and
$P^2 = 0$ with all P components being zero. The UIRs of the latter subclass 
have been analysed in \cite{shirokov}; they 
cannot have particle interpretation,
but a vacuum state may be represented by them.}), 
and the third class with $P^2 < 0$. Every class is
generated by a corresponding subgroup of the Poincar\'{e} group which is
called {\em a little group}. The little group describing the third class 
of representations is a group of rotations in 2+1 dimensions 
denoted by $O(2,1)$. The UIR's of this group were obtained and analysed 
a little bit later by V.~Bargmann \cite{bargmann}.  

In the 1960's Wigner returned to discuss the UIR's of the Poincar\'{e} 
group corresponding to particles with spacelike momenta \cite{wigner2}. He has
shown that quantum mechanical equations corresponding to these UIR's
describe particles with imaginary rest mass moving faster than light.
This almost coincided in time with the appearance of two seminal works in
which the hypothesis of faster-than-light particles was formulated explicitly, 
accompanied by a kinematic description of them \cite{bds} 
(see Appendix A) and by a quantum field theory of scalar tachyons\cite{fein}. 
The particles were called $tachyons$, from the Greek word 
$\tau \alpha \chi \iota \sigma$ meaning $swift$ \cite{fein}. 

These propositions immediately encountered strong objections related to 
the causality principle. It has been shown in several papers 
\cite{newton,roln,parment}, in agreement with an earlier remark by Einstein 
\cite{ein} (see also \cite{tolman,moller,bohm}), that by using tachyons as 
information carriers one can build a causal loop, making possible 
the information transfer to the past of an observer. This is deduced from 
the apparent ability of tachyons to move backward in time, which happens when
they have a negative energy provided by a suitable Lorentz transformation, 
this property of tachyons being a consequence of the spacelikeness of their 
four-momenta. A consensus was achieved that within the special relativity 
faster-than-light particles are incompatible with the principle of causality. 

Another important problem related to tachyons concerns the stability
of the tachyon vacuum. It is generally believed that any field theory
containing a negative mass-squared term in the Lagrangian has no stable
ground state (a vacuum) until the field is re-arranged converting tachyons
into ordinary particles with positive mass squared (for an instructive 
description of the problem   see e.g. \cite{nielsen}). This belief comes
from consideration of numerous models of spontaneous symmetry breaking,
the most famous being the Higgs model. Applied in a 
straightforward manner to consideration of faster-than-light particles
it results in a maximum (instead of a minimum) of the Hamiltonian for  
tachyon zero field, and leads to the conclusion that the 
existence of tachyons as free particles is not possible.

Fortunately, both problems turned out to be mutually connected and were 
resolved in the 1970's - 1980's.

The causality problem was resolved by the observation that fast tachyons, 
necessary to build the causal loop (so called transcendent tachyons), 
are not the objects of special relativity, being compatible, however,
with  general relativity \cite{pvcaus}. Then the situation with the causality 
violation by tachyons changes drastically since the modern cosmology 
based on general relativity supplies a preferred reference frame, 
so called comoving frame, in which e.g. the distribution of matter 
in the universe, as well the cosmic background (relic) radiation
are isotropic, and fast tachyons are extremely sensitive with respect to
this frame. In the preferred reference frame tachyons are ordered by
the retarded causality and have positively-defined energies. After the 
causal ordering being established in the preferred frame,
no causal loops appear in any other frame. 

Furthermore, in parallel with the 
causal ordering of the tachyon propagation one succeeds to get a stable tachyon
vacuum\footnote{It is possible to reverse this statement arguing that solving 
the problem of stability of the tachyon vacuum one ensures the causal behaviour
of tachyon fields.}, which presents the minimum of the field Hamiltonian
and appears, in the preferred frame, to be an ensemble 
of zero-energy, but finite-momentum, on-mass-shell tachyons propagating 
isotropically. Thus the space of the preferred frame is spanned by 
the continuous background of free, zero-energy tachyons; in some respects this 
is the reincarnation of the ether concept in its tachyonic version. 
    
We have to note that the stability of the tachyon vacuum distinguishes the
faster-than-light particles under consideration (i.e. genuine tachyons) 
from so called ``tachyons" appearing in numerous field-theoretical models 
with a negative mass-squared term in the Lagrangian mentioned above, 
in which these pseudo-tachyons have no stable vacua 
and therefore cannot be considered as particle-like objects, 
i.e. as elementary quantum field excitations above the minimum energy state of 
the field called vacuum, thus being condemned to disappear as faster-than-light
particles.     

This note is designed to present the solutions of the aforementioned problems 
of genuine tachyons and is organized as follows. In Section~2 the causal 
problem of the faster-than-light particles is described and a way to resolve 
it is shown. Section~3 addresses the problem of the instability of the tachyon 
vacuum, and a novel approach to the finding of that vacuum is suggested,
which turns out to be closely related to the causality problem solution.
Section~4 is devoted to some aspects of tachyon quantum field theories, 
including the unitarity problem and a non-locality of the tachyon states, which
leads, in particular, to a strong suppression of the effects of instability of 
ordinary particles with respect to the spontaneous emission of tachyons
considered in Section 5.
Section~6 describes the experimental status of the tachyon hypothesis.
Concluding remarks on a suggested approach to the tachyon problems 
and the note summary are presented in Sections~7,~8. 

In formulae used in this note the velocity of light $c$ and the Planck 
constant $\hbar$ are taken to be equal to 1.

\section{Causality problem of faster-than-light signals and its solution}
\setcounter{equation}{0}
\renewcommand{\theequation}{2.\arabic{equation}}

\subsection{Causality violation by tachyons in special relativity}
Let us present the causal paradox based on superluminal communications inspired
by the Tolman-M{\o}ller construction \cite{tolman,moller}. Our presentation 
of it is given in such a form of an explicit tachyon exchange that excludes any 
misinterpretations and wrong paradox solutions, e.g. those as given in numerous
publications by E. Recami and his coauthors (we refer to a few of them, 
\cite{recami}).
 
Consider a tachyon exchange between two observers $A$ and $B$ moving with 
respect to each other with 3-velocity $u$, each observer being equipped with 
an emitter and a detector of tachyons, with both emitters being able to emit 
tachyons having 3-velocities $v>1/u$ in an arbitrary direction. Let us call 
such tachyons which force the reinterpretation principle \cite{bds} to be 
applied to them (since both, the time intervals along their path and their 
energies change their signs when passing from the frame $A$ to the frame $B$ 
and vice versa) by transcendent tachyons\footnote{Thus transcendent tachyons 
are those which cross the boundary $E = 0$ under proper continuous Lorentz
transformations, see formula A.3 in Appendix A.}. Let us assume for 
simplicity that the detection of tachyons (and antitachyons) can be performed 
by time-of-flight (TOF) systems able to identify tachyons passing through 
them in any direction, with which the both observer's detectors are 
equipped. Let us call by a tachyonic event every passage of a tachyon or an 
antitachyon through the corresponding TOF system of any of the 
observers\footnote{Instead of a single tachyon (or antitachyon) one can assume 
a faster-than-light signal of an arbitrary complicated structure, 
e.g. modulated tachyonic beam.}.

The tachyon exchange is launched by the observer $A$ at the moment $t_0^A = 0$ 
(let us designate the observer's $A$ and $B$ times by respective superscripts) 
{\bf if and only if} no any tachyonic event has
been observed by him during a time interval of a length $T$ preceding  
$t_0^A$ (the necessity of this condition will soon become clear) 
\footnote{The value $T \geq D$, where $D$ is a distance between the 
observers at $t_0^A$ is sufficient for any set of the velocities $u$ and $v$ 
implemented in the process under consideration.} emitting at that moment 
a (transcendent) tachyon $\alpha$ to the observer $B$, as Fig.~1a illustrates.
   
The tachyon $\alpha$ reaches the observer $B$ and is detected by him at the
moment $t_1^A$. But in the frame of the observer $B$ this detection will look
like a passage of the antitachyon $\overline {\alpha}$ through the $B$ TOF 
system directed to $A$ at time $t_1^B$, the $t_1^B$ being earlier than 
$t_0^B$ since the tachyon $\alpha$ is a transcendent one (see Fig.~1b).
 
We insist (and shall prove) that such an ordering of the cause (tachyon 
emission at $t_0$) and its effect (the tachyon passage through the detector of 
the observer $B$), with the effect occurring before its cause 
as it is seen in the frame $B$, even being strange, does not mean, 
contrary to the general belief, any causality violation yet. 

Here we have to specify what is the causality and to define what is its 
violation.

\vspace{2cm}
\includegraphics[height=7.5cm,width=20.6cm]{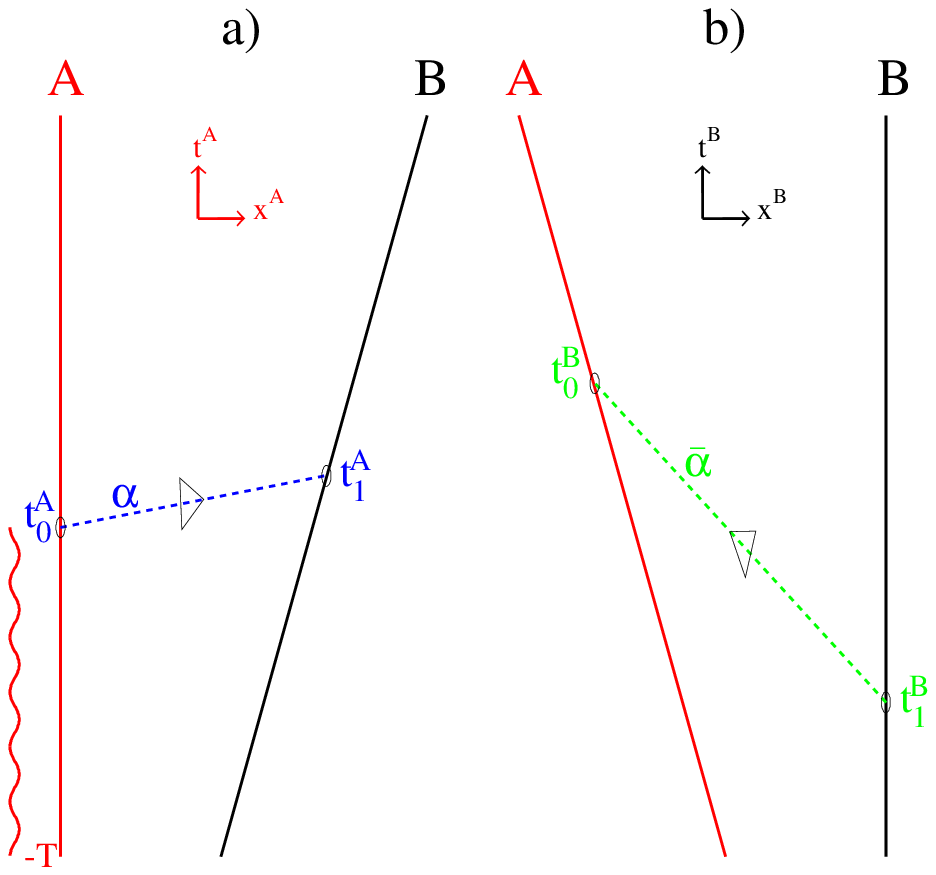}
Fig.~1. Exchange by a tachyon $\alpha$ as seen $a)$ by an observer $A$ to be 
an emission of this tachyon launched by him to reach an observer $B$, 
and $b)$ as seen by the observer $B$ to be a spontaneous emission of an 
antitachyon $\overline{\alpha}$ moving to the observer $A$. The observer's 
world lines are shown by solid lines, those for tachyons are shown by dashed 
ones. The ``dead time interval" preceding to the emission of the tachyon 
$\alpha$ at $t_0^A = 0$ is indicated by a wave line.
\vspace{5mm}

Of the many notions of the causal relations we select a single one, 
which can be formulated in two words: action$\rightarrow$ result. 
This, almost tautological definition, will be sufficient
for our analysis of the causal problem related with faster-than-light signals 
while makes its straightforward. Action and result (the cause and effect) are
related by a causal chain: by a beam of world lines of matter\footnote{Here 
we understand under the term matter any action carriers, e.g. photons or 
even gravitation if one includes consideration of Newton's apple.} 
which has no interruptions either in time or space. 
The existence of this material chain is an invariant thing: 
occurring in one reference frame it happens in any other frame. By definition, 
information flow goes from the cause to the effect, and this 
direction is also invariant whatever their time ordering could be (i.e. whether
we deal with a routine retarded causality or, e.g., with a hypothetical 
advanced causality in the spirit of Wheeler-Feynman absorption theory). 
The causal relations are governed by an extremely hard logical principle 
which is called {\em the causality principle:} any cause has an unalterable 
own origin. In physical language the causality principle is a requirement 
of the impossibility of creation of causal loops, i.e. the causal chains 
containing closed world lines.   

That is all. Nothing is required in the relevance of what the time ordering of 
a cause and its effect should be. For example, if the observer $B$ in our story,
having obtained the tachyon signal from the future (but from the space-like 
separated region), as shown in Fig.~1b, could not produce any influence to the 
signal sending (namely, to the emission of tachyon $\alpha$), i.e. the observer
$A$ would be inaccessible to the observer $B$, after detecting by the latter 
the tachyonic signal $\alpha$, during a whole interval T preceding the $t_0^A$,
then no problem with causality would appear.

Unfortunately, this is not the case for the story under consideration. The 
observer $B$ in our example has an access to the observer $A$ 
during the above interval since he has a tachyon
emitter equivalent to that of the observer $A$. So, at the time $t_2^B$ (see
Fig.~2b) he sends a faster-than-light signal (tachyon $\beta$) towards the 
observer $A$. If the velocity of the tachyon $\beta$ in the $B$ frame is higher
than that of the antitachyon $\overline \alpha$ the trajectory of the tachyon 
$\beta$ intersects the trajectory of the antitachyon $\overline \alpha$ 
somewhere in the space between the $B$ and $A$, and then tachyon $\beta$ will
reach the observer $A$ and will be detected by him at the time $t_3$ which, 
in the both frames, precedes the time  $t_0$. One can see that the possibility
of a causal loop is realized.
      
\vspace{2cm}
\includegraphics[height=7.5cm,width=20.6cm]{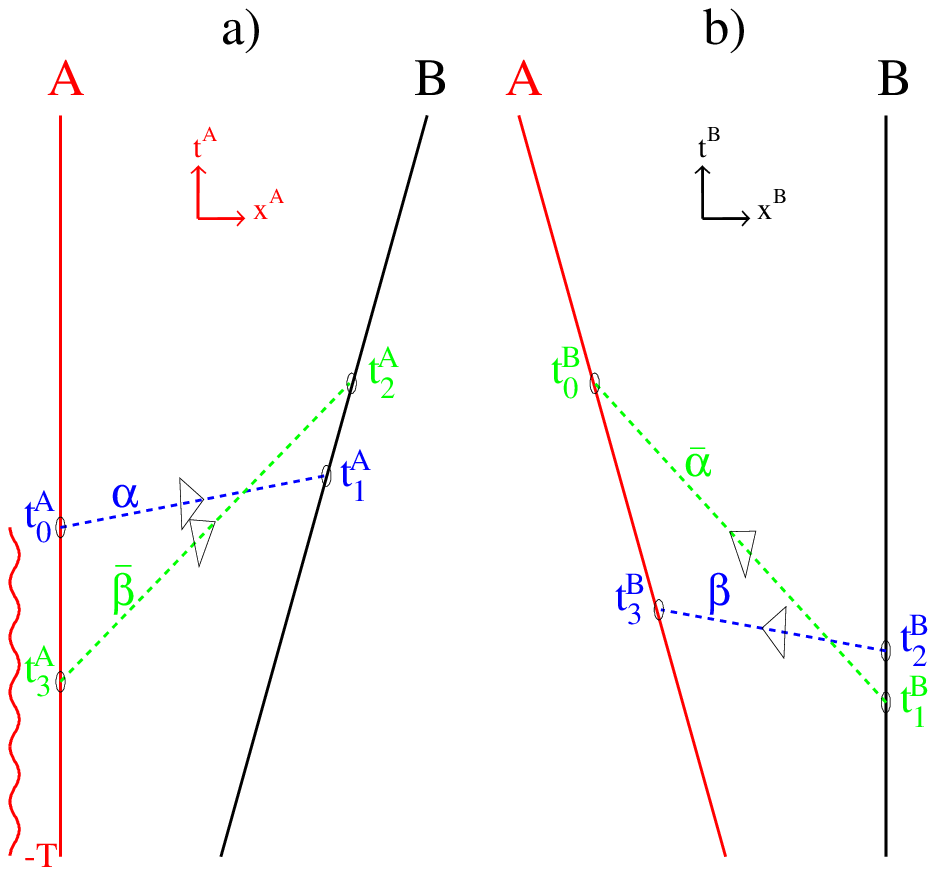}
Fig.~2. A causal loop (the Tolman-M{\o}ller paradox) as seen $a)$ 
in the frame of the observer $A$ and $b)$ in the frame of the observer $B$. 
\vspace{5mm}

Let us consider it in the frame of the observer $A$ (Fig.~2a). He will detect 
a tachyonic event (passage of an antitachyon $\overline \beta$ through its TOF
system\footnote{It is easy to prove that tachyon $\beta$ which is faster (in 
the frame $B$) than the transcendent antitachyon $\overline \alpha$, also is a 
transcendent tachyon, i.e. it changes the sign of its energy when transformed 
to the frame $A$.}) at the time $t_3^A$, i.e. inside the time interval $T$ 
preceding the launching time $t_0^A$. But the mandatory condition for 
launching the tachyon $\alpha$ was the absence of any tachyonic event during 
that interval! We have an unsolvable logical paradox, which was the main reason
of a rejection of the possibility of faster-than-light signals during a 
century, beginning since Einstein's formulation of this rejection \cite{ein}. 

In fact, the base stone of the causal paradox constructed above is the 2nd 
postulate of the special relativity declaring the equivalence of all 
inertial frames, which in our example was realized via the possibility for all 
observers to have tachyon emitters able to emit tachyons having arbitrarily 
large velocities, in particular $v>1/u$, in an arbitrary direction. As we shall 
see in the next subsection, this base stone turns out to be rather vulnerable 
when dealing with transcendent tachyons. 

\vspace{2mm}
But before going to that, three principal statements can be formulated at 
the end of this subsection as its key points:
\begin{itemize}
 \item [1.] Causality is an invariant entity which relates an effect and its 
cause in a certain (invariant) way.
 \item [2.] There is no explicit causality violation if an effect precedes its 
cause in time if this ``wrong" ordering takes place in spacelike separated 
regions. However there is a possibility of the causality violation if the wrong
ordering takes places in timelike separated regions, e.g. when a causal loop 
can be constructed.  
 \item [3.] Within special relativity there is a clear method
to design a causal loop with tachyons (e.g. via the Tolman-M{\o}ller 
construction) which excludes, within this theory, any chance for the existence 
of such particles able to serve as carriers of faster-than-light signals. 
\end{itemize} 

\subsection{Tachyons in an expanding universe}
Meanwhile, a realistic approach to tachyons shows that they can avoid this 
prohibitive sentence, appearing to be faster-than-light objects not obeying 
the special relativity, respecting however the laws of the general 
relativity. In this subsection we shall demonstrate how modern cosmology based
on general relativity treats tachyons in the expanding universe and prove 
that the use of the transcendent tachyons for building the causal 
paradoxes within the special relativity approach is illegal.
  
There is a strict criterion defining the limit of the application of the
special relativity in a given analysis called the geodesic deviation criterion.
It is formulated as follows \cite{misner}.          

Separate a space-time region in which you can carry out the measurements of
trajectories of your test particles with an accuracy of $\nabla \xi$. Then
trace in this region a free motion of test particles (their geodesics) with 
initially parallel world lines. If the world lines remain parallel within 
$\nabla \xi$ for any direction of the particle motion, you can state that the 
space-time of your region is a Minkovskian one, i.e. a flat and static 
space-time of special relativity, at least with an accuracy down to  
$\nabla \xi$. If it is not the case, you can either use an accurate general 
relativity approach, or remain within special relativity with the measured 
inaccuracy regarded as admissible.

However a situation is possible when the initially parallel geodesics deviate
drastically. This means that you have encountered some singularity, either of
the space-time, or behaviour of your test particles. In both cases you have no
choice and are obliged to use the general relativity approach in your analysis.
       
Let us apply this criterion to the motion of faster-than-light test particles,
first considering them classically, i.e. associating with them spacelike 
geodesics. How do such geodesics behave in our expanding universe?  

For the first time this problem was addressed by Davies \cite{davies}; a
thorough analysis of it was done a little bit later by Narlikar and Sudarshan
\cite{narlik}. They have shown that tachyons in the expanding universe undergo
a cosmological red shift of the same type as the ordinary particles, so the
expression
\begin{equation}
    pa = const
\end{equation}
is valid. Here $p$ is a tachyon 3-momentum, and $a$ is the scale factor in the 
metrics of the expanding universe: using spherical coordinates a line element
$ds^2$ in the isotropic universe can be written as
\begin{equation}
ds^2=dt^2-\frac{dr^2}{1\pm\frac{r^2}{a^2}}-r^2(\sin^2\theta d\phi^2+d\theta^2),
\end{equation}
where signs $+$ and $-$ are for open and closed universes, respectively.  
The equality (2.1) is obtained solving the spacelike geodesics equations.
In general coordinate form they look as follows:
\begin{equation} 
\frac{d^2 x^\mu}{ds^2} + \Gamma^\mu_{\nu\rho} \frac{dx^\nu}{ds} \frac{dx^\rho}{ds} = 0, 
\end{equation}
where $\mu,\nu,\rho$ run over 0,1,2,3, and $\Gamma^\mu_{\nu\rho}$ are 
Christoffel symbols. 

Introducing the coordinate $\chi$ according to $r = a\sinh\chi$ and 
$r = a\sin\chi$ for open and closed universes, respectively, and using instead 
of the time a quantity $\eta$ defined by the relation
\begin{equation}
dt = a d \eta
\end{equation}
in the spirit of \cite{ll1}, we obtain for the line element the expression 
\begin{equation}
ds^2 =a^2(\eta)~[d\eta^2 -d\chi^2 -\sinh^2\chi(d\theta^2+\sin^2\theta d\phi^2)].
\end{equation} 
in the case of an open universe, with the term $\sinh^2\chi$ being replaced by
$\sin^2\chi$ in the case of a closed universe. From the symmetry arguments we
can restrict our analysis to the two-dimensional case:
\begin{equation}
ds^2 = a^2(\eta)~(d\eta^2 - d\chi^2),
\end{equation}
valid for both, the closed and the open universes, as well as for the flat one. 
Then we have for non-zero $\Gamma^\mu_{\nu\rho}$'s the following relations:
\begin{equation}
\Gamma^0_{00} = \Gamma^0_{11} = \Gamma^1_{10}=\Gamma^1_{01}=\frac{a^\prime}{a}~,
\end{equation}
where the prime denotes differentiation over $\eta$. Thus equations (2.3) 
reduce to
\begin{equation}
\frac{d^2 \eta}{ds^2} + 
\frac{a^\prime}{a}\Big[\Big(\frac{d\eta}{ds}\Big)^2 + 
                       \Big(\frac{d\chi}{ds}\Big)^2\Big] =0
\end{equation}
and
\begin{equation}
\frac{d^2 \chi}{ds^2} + 2\frac{a^\prime}{a} \frac{d\eta}{ds}\frac{d\chi}{ds}=0.
\end{equation}

The integration of (2.9) gives
\begin{equation}
\frac{d\chi}{ds} a^2 = const = -iA_d,
\end{equation}
where $A_d$ is a real constant since $ds$ is imaginary, which together with the 
definition of the tachyon 3-momentum
\begin{equation}
p = i\mu  \frac{a d\chi}{ds}
\end{equation} 
leads to (2.1).
     
With (2.10) equation (2.8) integrates to
\begin{equation}
\frac{d\eta}{ds} = \pm \frac{1}{a}\sqrt{1-\frac{A^2_d}{a^2}}.
\end{equation}
Defining the tachyon energy $E=i\mu \frac{a d\eta} {ds}$ we get, after choosing
the negative sign in (2.12),
\begin{equation}
E =  \mu\sqrt{\frac{A^2_d}{a^2} - 1}.
\end{equation}

In the expanding universe the parameter $a = a(\eta)$ increases monotonically, 
thus it follows from (2.1), (2.13) that the tachyon momentum and energy are 
decreasing with the universe expansion. This is a behaviour which looks very 
similar to that of ordinary particles, but in the case of tachyons it has a 
drastic distinction: sooner or later tachyon momenta decrease to the value of 
$p = \mu$ and will tend to decrease further. But for real tachyons their 
momenta cannot become below $\mu$. So, what happens with a tachyon when its 
momentum reaches the limit of $p = \mu$ at $\eta$ equal to, say, $\eta_d$?

Let us assume that our universe at early times was filled with relic tachyons
created during the Big Bang\footnote{If it would be the case, the tachyons
would dominate the early universe by tens of orders of magnitude over the 
ordinary massive particles due to an infinite number of polarisation states
available per tachyon as it follows from the conclusion deduced in Sect.~4 
that tachyons, if they exist, have to be realizations of the 
infinite-dimensional representations of the Poincar\'{e} group. The dominance
of tachyons in the early universe would lead to a cosmology model of a quite 
different type as compared to the standard (inflationary) cosmology. However 
the consideration of the tachyon cosmology lies out of scope of this note.}.
To make the picture symmetrical let us suppose that for every tachyon an 
antitachyon of the same energy moving in the opposite direction existed 
somewhere. Evolving accordingly to the law (2.1) all these tachyons, as
follows from the tachyon vacuum model derived in Sect.~3.2,
dissolve (disappear) in the vacuum at $\eta_d$.  

Here we have to invoke a bit of a quantum mechanical consideration of the 
tachyon behaviour in the expanding universe; for the first time it was given 
in \cite{narlik} and starts as follows.

The Klein-Gordon equation for tachyons in the expanding universe is solved. 
Its solutions at $\eta < \eta_d$ present functions oscillating in time, 
typically as $\exp(-iEt)$. But for $\eta > \eta_d$ the solutions become damped 
(which can be seen also from the fact that the tachyon energy (2.13) becomes 
imaginary at $a > A_d$, i.e. at $\eta > \eta_d$), with the characteristic 
damping time given by $(\mu^2 H_d)^{-\frac{1}{3}}$, where $H_d$ is the Hubble 
constant at $\eta_d$ (with $\mu = 1$~GeV the damping time today
would be $5 \times 10^{-11}$~s). As we shall see in Sect.~3.2, the 
transition of tachyons to the vacuum state (tachyon ``dissolution" 
in the vacuum) at the tachyon momenta reaching $\mu$ agrees well with 
the tachyon vacuum model obtained in that section. 

However, just at this point our interpretation of the tachyon behaviour in the 
expanding universe starts to differ essentially from that proposed by Narlikar 
and Sudarshan. They believe that the tachyon trajectory at $\eta = \eta_d$ 
bends back in time (with the tachyon energy becoming negative) and then
propagates in counter-time direction, so the whole process looks as an 
annihilation at $\eta_d$ of a pair of a tachyon and an antitachyon, 
both being produced previously in spacelike separated regions in a 
highly correlated manner needed to ensure the above annihilation. Evidently,
this immediately rises causal paradoxes, being incompatible also with our
treatment of the problem described below (Sect.~2.3), and leads to a conclusion
about inevitability of the tachyon dissolution in the vacuum at $\eta_d$. 

Let us return to the classical consideration of spacelike geodesics approaching
the limit of $p = \mu$. Assuming the possibility of tachyon production 
not only in the early universe but in the present epoch too, let
us guess what would be the length $R$ of a trajectory of a tachyon until 
it reaches the ``time" of damping $\eta_d$ (when its dissolution in the vacuum 
occurs), in dependence on the initial momentum $p_0$ of the tachyon produced 
at $\eta_0$.

In order to answer this question we need the explicit form of the
scale factor $a$: 
\begin{equation}
a = a_0 (\cosh\eta - 1), ~~~ a = a_0 (1-\cos\eta)
\end{equation}
for the open and closed universes. At small $\eta$ the relation
\begin{equation}
a = \frac{a_0}{2} \eta^2 
\end{equation}
approximates $a$ for both, the closed and open universes, as well as for the 
flat one. Assuming that this approximation is still valid at the present epoch
and noting that $a d\chi = dr$ in this case, we rewrite (2.10) as
\begin{equation}
-i A_d = a \frac{dr}{ds} = a \frac{dr}{d\eta} \frac{d\eta}{ds}
\end{equation}
and using (2.12) obtain
\begin{equation}
dr = \frac{a A_d}{\sqrt{A^2_d - a^2}} d\eta  
\end{equation}
 
From the tachyon momentum definition (2.11) and relation (2.10), and 
accounting for (2.1), it follows that
\begin{equation}
\mu A_d  = p_0 A_0 ,
\end{equation}
where $A_0$ is the scale factor $a$ at $\eta_0$.    
Substituting $A_d = A_0 \frac{p_0}{\mu}$ followed from (2.18)
to (2.17) we obtain for the length $R$ the expression:
\begin{equation}
R(p_0) =  \int_{\eta_0} ^{\eta_d} \frac{a A_d }{\sqrt{A^2_d - a^2}} d\eta =  
\frac{a_0\eta^3_0 p_0}{2\mu} ~\Biggl(\sqrt{\frac{p_0}{\mu}} S +
\sqrt{\frac{p_0 - \mu}{p_0 + \mu}} ~\Biggr)  
\end{equation}
since
\begin{equation}
\eta_d = \sqrt{\frac{2A_d}{a_0}} = \eta_0 \sqrt{\frac{p_0}{\mu}}.
\end{equation}
$S$ in (2.19) is a superposition of elliptic integrals which can be reduced to
\begin{equation}
S = \int_{\alpha_0} ^{\pi/2} \frac{\cos^2 \alpha d\alpha}
{\sqrt{1-\frac{1}{2}\sin^2 \alpha}},
\end{equation}
where $\alpha_0$ is $\sin^{-1}\sqrt{\frac{2\eta^2_0}{\eta^2_0 + \eta^2_d}}$.
The curve (2.19) is plotted in Fig.~3. The prominent feature of this curve is
the singularity of its derivative at $p_0 \rightarrow \mu$, that is at the
tachyon energy approaching zero (namely, developing a pole $\sim 1/E$), coming 
from the second term in expression (2.19), while the derivative of the first 
term in this expression smoothly vanishes at that moment.

\includegraphics[height=14cm,width=28cm]{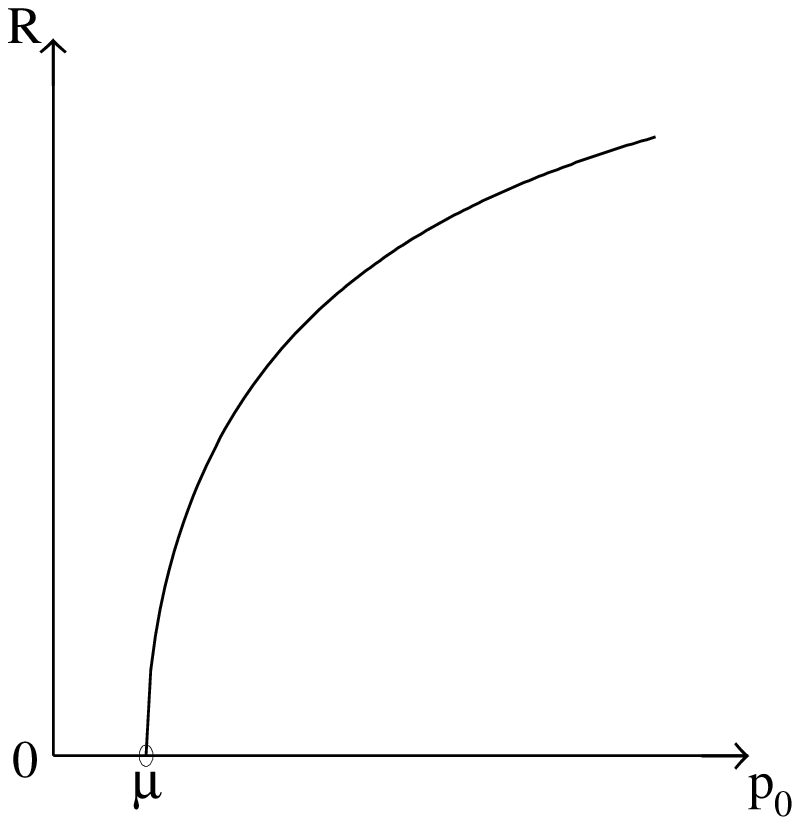}
Fig.~3. Tachyon trajectory length (from the production of a tachyon to its
``dissolution" in the vacuum) in the expanding universe $vs$ the tachyon 
initial momentum $p_0$. 

\vspace{5mm}
This means that the spacelike geodesics starting with close $p_0$'s at $\eta_0$
deviate strongly at the end of their trajectory.
According to the geodesics deviation criterion introduced at the beginning of
this subsection, this makes the zero-energy boundary for tachyons propagating in 
the expanding universe to be the ``no go" limit of the special relativity 
approach with a consequence that the transcendent tachyons, which have to cross
this boundary under the application of the Lorentz transformations, cannot be 
used as a legitimate material for building the causal paradoxes with 
faster-than-light signals. In the next subsection we consider the solution 
of these paradoxes respecting this ``no go" limit.

\subsection{Causal ordering of faster-than-light signals}
We have considered tachyon behaviour in the cosmological environment 
and have seen that fast tachyons must be treated with care since they are
very sensitive to the cosmological properties of the universe. Unfortunately,
an important role of the reference system in this treatment has been 
hidden; in order to unveil it we shall give two fragments of a text from the 
book by Landau and Lifshitz ``The Classical Theory of Fields", the fragments 
being devoted to the concept of a system of reference in general relativity 
and, in particular, in cosmology.
   
Such a system is not ``a set of bodies at rest relative to one another in
unchanging relative positions... for exact determination of the position of
a particle in space we must, strictly speaking, have an infinite number of
bodies which fill all the space like some sort of ``medium". Such a system
of bodies with arbitrary running clocks fixed on them constitutes a reference 
system in the general theory of relativity.

In connection with the arbitrariness of the choice of a reference system, the 
laws of nature must be written in the general relativity in a form which is
appropriate in any four-dimensional system of coordinates (or, as one says, in
$``covariant"$ form). This, of course, does not imply the physical
equivalence of all these reference systems (like the physical equivalence of 
all inertial reference systems in the special theory). On the contrary, the
specific appearances of physical phenomena, including the properties of the
motion of bodies, become different in all systems of reference." \cite{ll2} 

For example, when considering the phenomena related to cosmological effects 
their ``specific appearance" looks the most clear in the reference system 
connected to the universe as a whole, which becomes a natural preferred 
reference frame for this consideration. An instructive definition of this 
preferred frame is given in \cite{ll1}:

``The most convenient is a ``co-moving" reference system, moving, at each point
in space, along with matter located at that point. In other words, the 
reference system is just the matter filling the space; the velocity of the 
matter in this system is by definition zero everywhere. It is clear that this
reference system is reasonable for the isotropic model of the universe - for 
any other choice the direction of the velocity of the matter would lead to an 
apparent nonequivalence of different directions in space...

In view of the complete equivalence of all directions, the components 
$g_{0\alpha}$ of the metric tensor are equal to zero in the reference system
we have chosen. Namely, the three components $g_{0\alpha}$ can be considered
as the components of a three-dimensional vector which, if it were different 
from zero, would lead to a nonequivalence of different directions. Thus $ds^2$ 
must have the form $ds^2 = g_{00} (dx^0)^2 - dl^2$."

Just such a form have linear elements (2.2), (2.5), (2.6) which were used 
in our analysis of the spacelike geodesics ``appearance", i.e. the analysis was
carried out in the preferred reference frame defined above. To fix labels let 
us prescribe it to the system of the observer A in Sect.~2.1. 
Had we chose for the analysis the coordinate frame B moving with respect to the 
preferred one with the velocity $u$ then, accordingly to \cite{ll1,ll2}, 
a term proportional to $2u~dt~dr$ would appear in (2.2), a term proportional to 
$2u~a^2~d\eta~d\chi$ in (2.5), etc.

  
Our example of spacelike geodesics terminating their trajectories in the 
expanding universe rather dramatically shows that the mathematical apparatus 
of general relativity has to be applied in order to manage fast tachyons. 
This does not mean however that tachyons should be treated always on an equal 
footing with such objects of general relativity as black holes, clusters 
of galaxies, etc. But at least one step beyond special relativity has to be
done when dealing with tachyons: we have to abandon the second postulate
of special relativity declaring the equivalence of all inertial frames 
and replace it by a postulate of the existence of a preferred reference 
frame, to be a frame in which tachyon energy cannot become negative.

This means that reaching the zero energy level via any energy-decreasing 
process tachyons cannot tend to a further energy decrease, but must disappear 
at $E = 0$ due to a damping of their wave functions when going beyond this 
limit as has been noted above. Then no causal paradoxes involving tachyons 
can be built in the preferred reference frame since no material for their 
building (negative energy, counter-time tachyons) exists there\footnote
{In frames moving with respect to the preferred one the tachyon damping 
can occur at nonzero energies due to appearance of terms $g_{0\alpha}$ in the 
metrics, at both, positive and negative tachyon energies. Nevertheless, 
as will be shown below, no causal loops can appear in the moving frames too.}.  
      
However, there is a statement in the literature that the introduction of the 
preferred reference frame does not save the tachyon hypothesis from the 
causality violation in the expanding universe \cite{walstad}. Two observers, 
both residing in the preferred frame but separated by some cosmologically 
significant distances, move one relative to another in the expanding universe. 
As stated in \cite{walstad} this is the precise situation that leads to 
causality paradoxes: by means of sufficiently fast tachyons the observers can 
conspire to send messages into their past.

This argument becomes wrong in our approach. As illustrates Fig.~3 the maximal
distance passed by tachyons having the initial momenta $p_0$ in the expanding 
universe is $R(p_0)$. If it is less than the separation between the observers 
the tachyons under consideration never reach the opposed observer\footnote{
We have to note that the situation suggested by Walstad, i.e. an 
arrival of tachyons going back in time to cosmologically separated observers 
could indeed occur if one adopts the proposition by Narlikar and Sudarshan 
postulating the bending of the tachyon trajectories back in time as mentioned 
in Sect~2.2.}; if it is greater, the arrived tachyons will still have a 
positive energy, supplying no material for the building of a causal paradox. 
Trivially, this is a direct consequence of the definition of a preferred 
reference frame as a frame in which no negative energy tachyons can appear.  

Thus we may emphasize once more the clue property of the preferred reference 
frame: it is the frame in which the cause and effect are ordered by retarded 
causality. In other frames this may not be valid if tachyons do exist, 
however no casual loops can appear in them. This can be easily proved by noting
that if a causal loop has appeared in some non-preferred frame, this would mean
its appearance in the preferred frame which contradicts the definition 
of the latter introduced above.
This follows from the fact that any causal loop contains a timelike piece
of the world line, on the edges of which a cause and effect are ordered
in a wrong way (the effect precedes the cause). Since no proper
coordinate transformation can change the sign of a time interval along a 
timelike world line this wrong ordering would be conserved in the preferred 
reference frame too.

Now we have to make an important remark.

The postulate of the preferred reference frame, induced by a 
general-relativistic consideration of tachyons, can be accommodated easily in 
a flat and static space-time, i.e. in the Minkowski space-time, thus  
conserving the viability of the Lorentz group. Really, the derivation of the 
Lorentz transformations is based on the requirement of invariance of the 
interval (a line element in four-dimensional space-time, e.g. in the form
$ds^2 = dt^2-dx^2-dy^2-dz^2$) when passing from one inertial frame 
to another \cite{ll3}. 
The presence or the absence of the preferred reference frame among frames 
under consideration does not affect the derivation in any extent, while the
introduction of the preferred frame to the Minkowski space can be considered 
as an asymptotic influence of the real world on that space. In view of
this possibility to retain the Lorentz group when treating tachyons, free 
of causal violations, all the considerations which follow below will be 
restricted to a flat and static space-time, nevertheless with the preferred 
reference frame being involved in them to provide the above freedom.

First, one can find, within a quite general approach, a law 
ensuring the causality conservation independently on what is a time 
entanglement of causes and effects in various reference frames.

This law has to keep a causal order of the signal propagation along the causal
chain, i.e. to control the transfer of the matter throughout the chain.
Due to invariant properties of the causality it must be formulated in a 
covariant form, its mathematical expression having to be a scalar relation
which governs the signal propagation.  

Within the special relativity we have a single four-vector which can be used
for the aim of the causal ordering of signals, namely, the 4-momentum of
signal carriers $P$, and the causal condition can be written as 
\begin{equation}
P^2 \geq 0,
\end{equation}
which is equivalent to the statement ``cause always precedes effect in time"
expressed mathematically. Thus we can see again that within special relativity 
no faster-than-light signals are admissible. 

But the postulate of the preferred reference frame supplies us with
the second four-vector which can be used in the causal ordering formula, 
the four-velocity of the preferred reference frame with respect to an observer,
$U$ ($U = (1,{\bf u})/\sqrt{1-{\bf u}^2}$), so that formula can be written 
as follows:
\begin{equation}
(PU) \geq 0.
\end{equation}
In three-vector form this {\em boundary condition} leads to the restriction 
\begin{equation}
{\bf (vu)} \leq 1.
\end{equation}
This is just the condition necessary to destroy the causal loops 
described above. Thus with (2.23) the causality can be ensured, while the 
faster-than-light signals appear to be allowed (how the formula (2.24) works
will be explained in Sect.~3.2).

It is interesting to consider a particular case of expression (2.23), the
equality
\begin{equation}
(PU) = 0.
\end{equation}
In the preferred reference frame the four-vector $U = (1,0,0,0)$. 
Thus, in this frame the four-vector $P$ satisfying (2.25) cannot have non-zero
time component, i.e. it corresponds to a zero-energy tachyon propagating at 
infinite speed. As we have postulated above and shall prove later (Sect~3.2) 
the condition (2.25) is indeed a gauge fixing the tachyon vacuum.

Also it will be shown in Sect.~4.1 that the abandoning of the principle
of the equivalence of all inertial frames can be done without any damage 
for the description of physical processes by the Standard Model, 
if only the ordinary particles are involved in them. 
This principle simply has to be replaced by a requirement of a relativistically
covariant description of all physical processes, allowing to include those with
the interacting tachyons. Furthermore, it may be postulated that a theory of
such tachyons, being formulated in the preferred reference frame in which all
the processes (including both, tachyons and ordinary particles) are ordered by
the retarded causality, has to look as similar to the standard one as possible,
thus constituting the respective correspondence principle.

\section{Problem of tachyon vacuum instability and its solution}
\setcounter{equation}{0}
\renewcommand{\theequation}{3.\arabic{equation}}
It is quite surprising that the introduction of a preferred reference frame 
into the tachyon hypothesis helps to solve the next serious problem of it,
the problem of the instability of the tachyon vacuum which will be considered
in this section. 

\subsection{Standard treatment of the problem}
Consider, as a toy model, a free non-Hermitian scalar tachyon field with the 
Lorentz-invariant Lagrangian:
\begin{equation}
L = \int d^3 {\bf x} \Big{[}\dot\Phi^*(x)\dot\Phi(x) - 
\nabla\Phi^*(x) \nabla\Phi(x) + \mu^2 \Phi^*(x) \Phi(x)\Big{]}  
\end{equation}
The Hamiltonian of this field obtained as usual
\begin{equation}
 H = - L + \dot\Phi \frac{\partial L}{\partial \dot\Phi}            
\end{equation}
reads
\begin{equation}
H = \int d^3 {\bf x} \Big{[}\dot\Phi^*(x)\dot\Phi(x) +
\nabla\Phi^*(x) \nabla\Phi(x) - \mu^2 \Phi^*(x) \Phi(x)\Big{]}.  
\end{equation}
Now, if a finite value of the $\Phi$ which minimizes the Hamiltonian can be 
found, it will represent the ground state of the field called vacuum. 

A standard approach to finding the minimum of the Hamiltonian (3.3), 
$\delta H = 0$, is reduced to an analysis of the potential term of (3.3),
which assumes, implicitly, that the search for the ground state of the 
Hamiltonian is replaced by looking for its minimum under restrictions 
conditioned by the Lorentz-invariant pair of the vacuum ``initial" conditions: 
\begin{equation}
\Phi = const~in~time,
\end{equation}
\begin{equation}
\Phi = uniform~(const)~in~space,
\end{equation}
which is an a-priori hopeless exercise since the potential term of (3.3),
$V = \int d^3 {\bf x} (-\mu^2) \Phi^* \Phi$,
has a maximum at $\Phi = 0$ instead of the necessary minimum. 
In the case under consideration (a free tachyon field) 
this is interpreted as an impossibility (instability) 
of the tachyon vacuum, while in the models of of the spontaneous symmetry 
breaking containing additional term(s) in the Lagrangian, say 
$\lambda \Phi^4$, this results in a transition of the field to the true 
ground state accompanied, after the re-arrangement of the field accordingly to
the new vacuum, by disappearance of the initial tachyon converted to 
an ordinary particle.  

This mechanism, so loved by theoreticians, works perfectly in models of the 
spontaneous symmetry breaking, but it is not applicable to the 
faster-than-light particles.

\subsection{Stability of the tachyon vacuum}
As we have seen, the hypothesis of the faster-than-light particles requires 
consideration of tachyons under a postulate of a preferred reference frame
which is necessary for the causal ordering of the signals propagating over 
the spacelike intervals. This requirement must be respected by the procedure 
of the tachyon ground state finding also. Therefore the Lorentz-invariant pair 
of the initial conditions (3.4), (3.5) must be replaced by a single, 
Lorentz-non-invariant one:   
\begin{equation}
\Phi = const~in~time
\end{equation}
which separates, obviously, the preferred reference frame. Then the Hamiltonian
which has to be analysed in the search for the ground state will contain, 
together with the potential term, a gradient energy term to be read 
\begin{equation}
H = \int d^3 {\bf x} \Big{[} \nabla\Phi^*({\bf x}) \nabla\Phi({\bf x}) - 
\mu^2 \Phi^*({\bf x}) \Phi({\bf x})\Big{]}.  
\end{equation}
One can easily obtain the solutions of the equation $\delta H = 0$ directly,
but from the pedagogical point of view it is worth to go by a slightly
longer path noting that
our task to find the variation $\delta H = 0$ coincides now (see 3.2) with
the finding the variation $\delta L = 0$, which is
just the exercise of finding the (Euler - Lagrange) equation of motion.
For the scalar field of the Lagrangian (3.1) it is the Klein-Gordon equation
with the negative mass-squared term $-m^2 = \mu^2$: 
\begin{equation}
\Big{(}\frac{\partial ^2}{\partial t^2} - \partial _i \partial ^i - \mu^2\Big{)} \Phi(x) = 0~, ~~~~~~~~i = 1,2,3
\end{equation}
which has well-known solutions in the form of plane waves 
$\exp{ \pm i(Et - {\bf px})}$ with the dispersion relation \cite{bds,fein} 
\begin{equation}
E = \sqrt{\bf{p}^2 - \mu^2},~~~~~|\bf{p}| \geq \mu .
\end{equation}

Now we have to require the condition (3.6) to be fulfilled, and this can be 
easily satisfied by putting $E =0 $ in the obtained solutions, which 
evidently minimizes the Hamiltonian density in (3.3) to zero value. If, for
example, one ``pumps" in some way (via interactions) the energy into particular
vacuum modes promoting their conversion to real tachyons (field excitations), 
the field energy H will be increased correspondingly.  

Then the final result of our exercise of finding the tachyon vacuum state 
(in the preferred reference frame) can be presented as a superposition 
of plane waves corresponding to mass-shell, zero-energy tachyons 
propagating isotropically.
On the first glance the superposition may appear as the coherent sum of vacuum 
modes \footnote{At the moment (before second quantization) we treat the 
tachyon vacuum as a quasi-classical ground state (while performing second
quantization we will deal with the tachyon field excitations mentioned 
in the previous paragraph).} 
with opposite signs in exponentials,
\begin{equation}
tachyon~ground~state
~~\sim \sum_{all~directions} 
\big{[}\exp{i{\bf(p_0 x)}} - \exp{i{\bf(-p_0 x)}} \big{]}~,
\end{equation}    
where ${\bf p_0}$ is a vacuum tachyon momentum, $|{\bf p_0}| = \mu$, while 
a demand of the translational invariance of the vacuum state 
leads to the inclusion of the vacuum modes with
$\exp{i{\bf (p_0 x)}}$ and $\exp{-i{\bf (p_0 x)}}$ in
the coherent superposition with the phase shift equal to $\pi$.

However, being on the mass shell, i.e. having infinite velocities, the vacuum 
tachyons immediately escape the place of their appearance, being replaced
by other, newly born coherent tachyon-antitachyon vacuum pairs 
and by vacuum tachyons coming from a remote environment. 
The result is that the tachyon 
vacuum wave function has to be presented by an incoherent (stochastic) 
superposition of the tachyon vacuum modes which can be written as
\begin{equation}
tachyon~ground~state
\sim 
\int \Bigl[\exp[i({\bf p_0 x}  + \xi_{\theta\phi})] - 
           \exp[-i({\bf p_0 x} + \xi_{\theta\phi})]\Bigr] d\Omega~,
\end{equation} 
where $d\Omega=d\cos\theta d\phi$ is the element of the solid angle, 
and phases $\xi_{\theta\phi}$
are independent, uniformly distributed random variables 
in the interval $[0, 2\pi]$.
The average of this expression over the phases $\xi_{\theta\phi}$
(i.e. the expectation value of the vacuum field) vanishes, similarly to
(3.10), indicating that the tachyon vacuum is a stationary state
of the field, with {\em statistical fluctuations} of the vacuum modes,
entering the integral in (3.7). 

Schematically, the tachyon vacuum in the preferred reference frame 
can be presented by a line at $E = 0$ separating hatched and non-hatched 
regions of tachyon energy, see Fig.~4a. The latter region is the region 
allowed for real tachyons, i.e. for tachyons having non-zero energy.

\includegraphics[height=14cm,width=28cm]{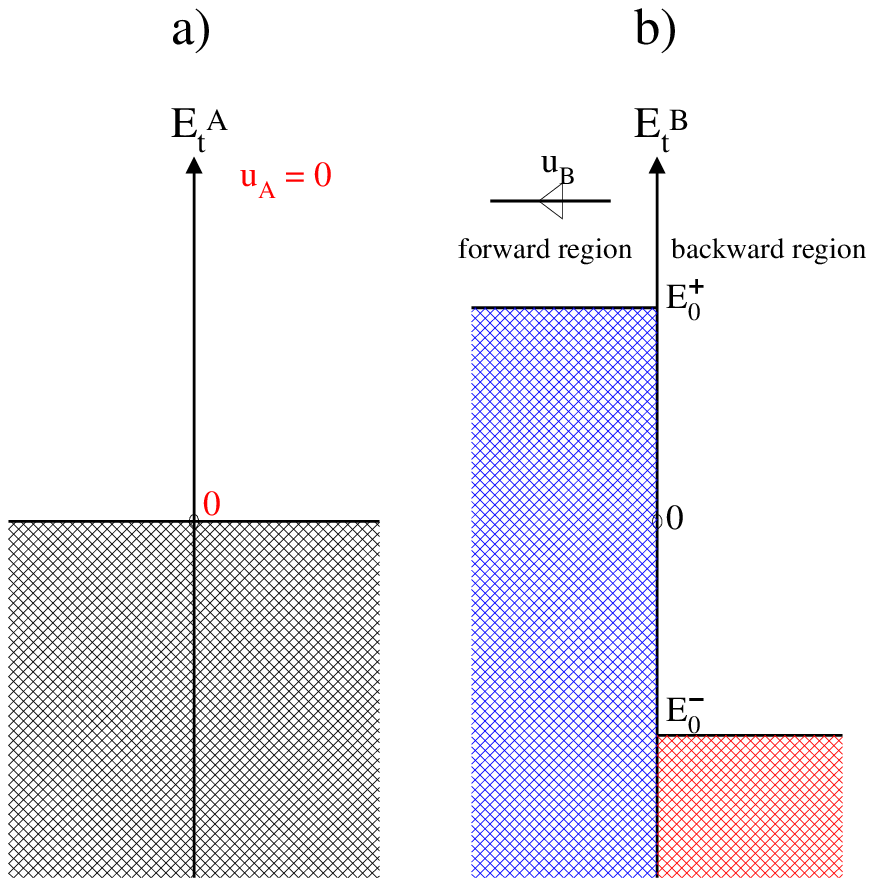}
Fig.~4. The tachyon vacuum energy levels as seen $a)$ from the preferred 
reference frame and $b)$ from a frame moving with respect to the preferred one
with 3-velocity~$u$. The direction of the preferred frame motion as seen from
the (moving) frame B is indicated by an arrow in the top part
of $b)$. $E_{0}^{+}$ and $E_{0}^{-}$ mark the ``forward" and the ``backward" 
tachyon vacuum energy levels in the moving frame given by (3.12) and (3.13). 
The vertical axes on both figures are for tachyon energies, with the hatched 
regions to be excluded domains for a proper (without reinterpretation)
emission and absorption of tachyons.
\vspace*{0.6cm}

It looks like we have obtained a paradoxical result: the tachyon vacuum, filled 
with on-mass-shell tachyons, has the averaged vacuum four-momentum (0,0,0,0), 
as it should be for the Lorentz-invariant vacuum.
However, being on the mass shell, the vacuum tachyons will acquire non-zero 
energies (positive and negative) when we pass to a frame moving with respect 
to the preferred one. In other words, the tachyon vacuum 
in the moving frames is asymmetric (i.e. violating the rotational symmetry), 
as illustrates Fig.~4b: in the direction coinciding with that of the preferred
frame velocity the tachyon vacuum boundary will have a positive value: 
\begin{equation}
E_0^+ = \frac{\mu u}{\sqrt{1 - u^2}},
\end{equation}  
where $u$ is 3-velocity of the preferred frame with respect to the observer,
while in the opposite direction the tachyon vacuum boundary will be shifted to 
the negative value:
\begin{equation}
E_0^- = -\frac{\mu u}{\sqrt{1 - u^2}}.
\end{equation}
Just the boundary (3.12) will prevent the observer $B$ of Sects.~2.1, 2.3 
from sending the causality-violating signal to the observer 
$A$ since the maximal tachyon velocity in that direction allowed to him, 
as can be seen from (3.12) and dispersion relation (3.9), is $1/u$. Thus 
one can say that all acausal tachyon states get confined in the tachyon vacuum.

In the next section we shall suggest a covariant prescription for such a 
confinement. But here we would like to emphasize an important property of 
the tachyon vacuum built in such a manner. Namely, it is invariant,
in the following sense: if no real tachyons exist in the preferred reference
frame, no such tachyons exist in any other (inertial) frame, and vice versa.  
   
In order to conclude this section, let us compactify all the considerations 
presented in it via the Lagrange formalism. It can be done modifying the
Lagrangian (3.1) by adding to it a Lorentz-non-invariant term proportional to
the 4-velocity of the preferred reference frame:
\begin{equation}
L = \int d^3 {\bf x} \Big{[}\dot\Phi^*(x)\dot\Phi(x) - 
\nabla\Phi^*(x) \nabla\Phi(x) + \mu^2 \Phi^*(x) \Phi(x) -
i \lambda U^\mu \partial_\mu [\Phi(x) - \Phi^*(x)] \Big{]},  
\end{equation}
where $\lambda$ has the dimensionality of the mass squared. Note that the 
additional term does not change the equation of motion, so all the results
obtained with (3.8) remain valid. Choosing  $\lambda = \mu^2$ one gets 
the corresponding Hamiltonian
\begin{equation}
H = \int d^3 {\bf x} \Big{[}\dot\Phi^*(x)\dot\Phi(x) +
\nabla\Phi^*(x) \nabla\Phi(x) - \mu^2 \Phi^*(x) \Phi(x) -
\frac{i \mu^2{\bf u}}{\sqrt{1-u^2}} \nabla [\Phi(x) - \Phi^*(x)]\Big{]}.  
\end{equation}
Thus, the additional term in the integrand of (3.15) shifts the tachyon vacuum 
energy boundaries accordingly to (3.12), (3.13), depending on the direction of 
the 3-velocity of the preferred reference frame ${\bf u}$. 

The fact that the Lorentz-non-invariant term in the Lagrangian (3.14) does not
change the tachyon equation of motion has several important consequences. 
A tachyon Feynman propagator, which is, similarly to propagators of ordinary 
particles, a Green's function for the tachyon equation of motion, will not 
acquire any Lorentz-non-invariant admixture, and this means that the 
Lorentz-non-invariance will be restricted, in our model, to the 
asymptotic-tachyon-states sector only. The importance of this conclusion can be
stressed by the result, followed from the fact of the Lorentz-invariance of 
the tachyon Feynman propagator (which together with the Lorentz-invariance of 
the Feynman propagators of all other particles, 
maintained in our model of the Lorentz-invariance violation), that the speed of
light remains an unique, universal velocity constant which limits particle 
velocities at both sides of the light barrier. In particular, an explicit 
breaking of the Lorentz symmetry by adding to the Lagrangian 
the Lorentz-violating term which affects the particle propagators, considered 
by Coleman and Glashow \cite{colgla1,colgla2}, among many others
(which leads to individual maximum attainable velocity for each fundamental 
field, differing from the velocity of light) is not relevant to our approach. 
For the same reason the strong restrictions on multiple Lorentz-violating 
coefficients compiled in the ``Data Tables for Lorentz and CPT violation" 
\cite{kostel} are not applicable to our considerations. 

Note, a thorough review of models violating Lorentz invariance is 
given in \cite{mattingly}. It contains 281 references. Unfortunately, 
none of them is relevant to our approach.
    
\section{Quantum aspects of the tachyon theory which are to be revised}
\setcounter{equation}{0}
\renewcommand{\theequation}{4.\arabic{equation}}
\subsection{Confinement of acausal tachyons}
Quantum field theories of tachyons like that by G. Feinberg \cite{fein},
can be easily modified to include the boundary condition (2.23) 
and the tachyon vacuum gauge (2.25) in the tachyon field operators. 
So, Feinberg's expression for a scalar tachyon field operator
(expression (4.1) in \cite{fein}) can be modified as follows:  
\begin{equation}
\Phi(x) = \frac{1}{\sqrt{(2\pi)^3}} 
\int{d^4k~\Big{[}a(k)\exp{(-ikx)} + 
a^+(k)\exp{(ikx)}\big{]}~\delta(k^2+\mu^2)~\Theta(kU)},
\end{equation}
where $k$ is a tachyon four-momentum, $a(k), a^+(k)$ are annihilation and
creation operators with bosonic commutation rules, annihilating or creating 
tachyonic states with 4-momentum $k$, and $U$ is a four-velocity of the 
preferred reference frame with respect to (any particular) frame in which 
the tachyon field quantization is carried out. 

One can see that the expression (4.1) is explicitly Lorentz-covariant.
This covariance includes the invariant meaning of the creation and annihilation
operators defined in the preferred frame;
thus, for example, an annihilation operator $a(k)$ remains an
annihilation operator $a(k^\prime)$ in the boosted frame, even if the zero
component of $k^\prime$ may become negative. This is because the one-sheeted 
tachyon mass-shell hyperboloid is divided by the covariant boundary 
$\Theta(kU)$ into two parts separated in an invariant way. In particular,
the condition (2.25) results in a possibility of the standard operator
definition of the invariant vacuum state $|0>$ via the annihilation operators 
$a(k)$, $a(k)|0> = 0$ for all $k$ such that $|{\bf k}| > \mu$, 
because the field Hamiltonian turns out
to be bounded from below (see formula (C.13) in Appendix C where toy models 
of scalar tachyon fields are considered), and of the construction of the 
invariant Fock space as usual:
\begin{equation}
 |k_1,k_2,...,k_i,...> = a^+(k_1)a^+(k_2)...a^+(k_i)...|0>.
\end{equation} 

Since now the Lorentz boosts do not mix creation and annihilation operators
one is not forced to use anticommutator relations for 
scalar tachyon fields as it was necessary in Feinberg's theory. 
The commutation relations maximally close to the canonical ones may be used
(see Appendix C), and this leads to other distinctions of our approach
as compared to Feinberg's theory: to a local Hamiltonian of the scalar 
tachyon field and to a possibility of use of the Lagrange formalism 
in a construction of the model lost in his theory. 

When calculating the tachyon production probabilities and cross-sections 
the confining $\Theta$-functions will accompany the production amplitudes as 
factors restricting the reaction phase space, so the expressions for
these probabilities can be displayed as follows:
\begin{equation}
W = \int| M |^2 d\tau \prod_{i} \Theta(P_i U),
\end{equation}
where M is a matrix element of the reaction (which has to be representable in a
Lorentz-invariant form), $d\tau$ is a reaction phase space element, and the 
product of $\Theta$ functions includes all the tachyons 
(having 4-momenta $P_i$) participating in the reaction. 

For the first time this formula was suggested and exploited in a paper 
\cite{pvmich}, which was dedicated to an
experimental search for the tachyon preferred reference frame using the Earth 
motion \cite{pdg} with respect to this frame\footnote{Motivation for formula 
(4.3), based on the causality problem solution, which led to this experiment,
was given in \cite{pvannih}.}.

\subsection{Towards the unitarity of a tachyon theory}
The replacement of the second postulate of special relativity, requiring 
the equivalence of the inertial frames, by a softer demand for physical
processes to be described in a covariant manner cures many diseases of
the Lorentz-invariant tachyon theory. In addition to the solution of the 
causality problem and the problem of the tachyon vacuum instability related to 
it, the introduction of the preferred reference frame into the physics of 
interacting tachyons removes difficulties related to the unitarity of such 
interactions.

In particular, the unitarity paradox was built in the reference \cite{unita}.
Though it was related to a particular theory of tachyons \cite{as,ds} it can be 
generalized to work against any theory of interacting tachyons unless the
tachyon vacuum is fixed by the gauge (2.25) resulting in the appearance 
of restricting $\Theta$-terms in (4.3). With these terms the paradox 
will be destroyed since meson decay to itself and to a tachyon, considered 
in \cite{unita}, will have zero probability if the meson is at rest in the 
preferred reference frame due to a corresponding term $\Theta(PU)$ in
(4.3) forbidding production of negative-energy tachyons in the preferred frame 
(similarly to a prohibition of an analogous decay with the tachyon replaced by 
an ordinary particle of negative energy, 
with the corresponding term $\Theta(E)$ in the decay phase space). 
This will be true in any frame with the reaction kinematic 
parameters corresponding to the meson at rest in the preferred frame. On the 
other hand, in a frame moving with respect to the preferred one the process of 
such (spontaneous) decay can be allowed (if the meson appears in flight in the 
preferred frame), validating the formal calculations carried out in 
\cite{unita}. Some examples of processes of this type and comments on them 
are presented below, in Sect.~5.

In a similar way most of the arguments of paper \cite{unita2} leading to a 
conclusion about inconsistency of any theory containing 
interacting tachyons with unitarity can be invalidated. 

To generalize the application of the concept of the preferred reference 
frame to the unitarity problem consider, in this frame, a reaction with 
tachyons present in the initial and/or final states. If one assigns to all 
initial state tachyons labels ``cause", and to all final state tachyons labels 
``effect", this labelling can be easily traced to any reference frame due to 
its invariance ensured by causal terms in the product of formula (4.3),
even though the tachyons of the initial state in the preferred frame  
might appear, in some other frame, as apparent antitachyons 
in the final state of the reaction 
due to the reinterpretation principle application, and vice versa.
Then the tachyonic initial and final (asymptotically free) states will get the 
same status from the unitarity point of view as those of the ordinary particles.
In other words, this means that the reaction asymptotic ``in" and ``out" 
Fock spaces are unitarily equivalent, even when they include tachyonic Fock 
spaces (4.2), just due to the invariant meaning of the tachyon creation 
operators $a^+(k)$.

Perhaps, the full content of this subsection may be expressed in a few words,
namely: tachyon theories with the gauge (2.25) for the tachyon vacuum can 
be made unitary.

\subsection{Tachyons within the Poincar\'{e} group approach}
In spite of being very important for consistency of a tachyon theory the 
necessity of the preferred reference frame (needed to ensure the reasonable,
i.e. causal, vacuum-stable, unitarity-consistent behaviour of faster-than-light
particles) does not exclude the consideration of tachyons within the special 
relativity (i.e. the Poincar\'{e} group) approach. The causal restrictions on 
production and propagation of tachyons introduced in previous sections become 
experimentally essential for tachyons with velocities greater than $c^2/u$, 
where $u$ is the velocity of the observer with respect to the preferred frame. 
For Earth-based laboratories the velocity $u \approx 370~km/s$ \cite{pdg},
therefore such restrictions become important only for tachyons having velocities
exceeding that of light by a factor of about 800 (note that such tachyons are
rather elusive objects possessing very low energy losses and are
difficult to handle experimentally). 
For relativistic tachyons (having velocities close to~$c$) special 
relativity remains a very good approximation and their description by 
the Poincar\'{e} group UIR's is quite adequate.

Therefore it is instructive to consider tachyons as realizations 
of these representations.

\subsubsection{Tachyons as realizations of the infinite-dimensional unitary 
irreducible representations of the Poincar\'{e} group (discrete series)}
Several classes of UIR's of the little group $O(2,1)$ of the Poincar\'{e} group,
describing imaginary rest mass particles, were considered by Wigner in 
\cite{wigner2}: two classes of infinite-dimensional UIR's (continuous and 
discrete classes) and a class of trivial representations corresponding to 
spinless (scalar) tachyons.  

Many tachyon quantum field theories which exploited the latter class of
the UIR's, leading to the Klein-Gordon equation with a negative mass term,
can be found in the literature. However, from our point of view, the models of 
scalar tachyons have to be considered as toy models only. The argumentation 
for this is the following.

First, scalar tachyons interacting with ordinary particles would lead 
to an appearance of poles in the particle interaction amplitudes due to 
exchange by tachyons located on the mass shell
as was indicated originally by \cite{kostya} and then by \cite{unita2}. 

Furthermore, the existence of scalar tachyons would result in the instability 
of photons, either via photon decay to a tachyon-antitachyon pair, or via such 
a decay accompanied by a photon of a lower energy, i.e. via reactions:
\begin{equation}
   \gamma \rightarrow t \overline t
\end{equation} 
\begin{equation}
   \gamma \rightarrow \gamma^\prime t \overline t
\end{equation}
(one of these reactions has to be suppressed by the $C$-parity conservation, 
depending on the $C$-parity of the tachyon-antitachyon pair). Avoiding this 
photon instability leads to a conclusion about necessity of a very low coupling
of scalar tachyons to ordinary particles, including photons,
since apparently no other mechanisms exist to suppress these reactions in this
case. Analogous conclusion can be drawn for the Cherenkov radiation by tachyons
which is related to reaction (4.4) via crossing symmetry.

These are the reasons why we believe, together with \cite{camen} and contrary 
to the mainstream of the tachyon models suggested, that the most suitable 
representations to be related with tachyons are infinite-dimensional UIR's of 
the Poincar\'{e} group.
Though as noticed in \cite{camen}, a self-consistent and covariant 
field theory for these representations does not exist, besides the approaches 
proposed in \cite{ruhl,sudmuk,hama,barut}, the main characteristics of tachyons
considered as realizations of these UIR's can be deduced from general
properties of the little group $O(2,1)$ and its generators, briefly 
described in Appendix B. 

First, all the states of these UIR's possess non-zero values of $helicities$, 
which can run up to infinity since the representations are infinite-dimensional.
This can impose strong restrictions on the tachyon production amplitudes
resulting from the angular momentum conservation, which can explain, at least
partially, the failure of tachyon search experiments. 
Further, when choosing between continuous and discrete classes of the 
representations, the latter has an essential advantage since
the two branches of this class have an attractive property which
ensures the Lorentz-invariant separation of tachyon and antitachyon states. 
As noticed in \cite{camen}, the Pauli-Lubanski vector
({\em the spin operator}) defined by
\begin{equation}
W^\mu = \frac{1}{2} \epsilon^{\mu\nu\rho\sigma} P_\nu M_{\rho\sigma},
\end{equation} 
$\epsilon^{\mu\nu\rho\sigma}$ being the fully antisymmetric tensor, is timelike
for the discrete class and spacelike for the continuous one (the latter is true
also for the Pauli-Lubanski vector in the case of ordinary, $P^2 > 0$, 
particles). This means that the time component of this vector does not change 
its sign under Lorentz transformations in the former case\footnote{With the 
specific vector $P^0$ defining the little group (see B.1), 
chosen to be $(0,0,0,\mu)$, this component equals to $\mu M_{xy}$.},
and may change the sign under a suitable Lorentz transformation in the latter 
case. This property of the Pauli-Lubanski vector allows the invariant 
definition of tachyon and antitachyon states 
when choosing the discrete class for their description. 
This can be seen also from the fact that the representations
of two discrete series, $D_s^+$ and $D_s^-$, are complex conjugate, thus
corresponding to equal mass particles and antiparticles defined invariantly. 
In view of this invariance the reinterpretation principle
suggested in \cite{bds} and aimed at the reinterpretation of negative energy 
tachyons as antitachyons moving in the opposite direction in space and time
(Appendix A), appears to be restricted to the kinematic domain only (where
it can be quite useful when doing a kinematic analysis).
In other words, defined in such a manner particular numbers of 
faster-than-light particles and antiparticles participating in a given reaction 
remain Lorentz-invariant, independently on the fact 
that the signs of the time components of four-momenta of some tachyons 
and/or antitachyons, i.e. the signs of their energies, can be changed 
by Lorentz transformations. A direct consequence of this is the
invariance of a global vacuum state defined as a state without localized 
quantum excitations, i.e. real particles (thus if some volume restricted by 
a spacelike surface does not contain real particles in some reference frame, 
there are no real particles in that volume in any other inertial frame) 
\footnote{It may sound paradoxically, but the invariance of the tachyon
vacuum state is provided by the non-invariance of the tachyon vacuum energy
boundaries.}. It is worth to note here that all these features, 
including the invariance of the global vacuum state, would be impossible 
(within the Lorentz approach) for scalar tachyons, which is illustrated by 
considerations of particle aspects of scalar tachyon models suggested in 
Refs. \cite{fein,as,sud2,fein2}, according to which 
a zero- or single-tachyon state in one reference frame 
can be transformed by a Lorentz boost to the state with an infinite number 
of positive-energy tachyons in another frame,
which immediately rises a question about conservation of unitarity in these 
models.

The invariance of the number of real tachyons under Lorentz transformations 
completes the labelling of tachyons as possessing positively-defined energy 
in the preferred reference frame, suggested before. 
Being defined in this frame as belonging to either $D_s^+$ or
to $D_s^-$ branch, tachyons and antitachyons obtain fixed labels which are 
independent on whether the signs of their energies change or not when passing 
to an arbitrary reference frame. 

Our choice of $D_s^\pm$ representations means that tachyons should be described
by infinite-component wave equations \cite{ruhl,hama}. 
This, in turn, leads to a consequence that tachyons must be produced in pairs 
with antitachyons only, in order to ensure that their production amplitudes 
would be scalar functions (expressed in a Lorentz-covariant form). 
Thus {\em the tachyon number}, postulated as a difference
between numbers of tachyons and antitachyons, having essentially non-zero 
energies in the preferred reference frame, may be considered as a good,
conserving quantum number, at least until cosmological effects are 
taken into account \cite{pvcaus}. 

As to the problem of the tachyon vacuum stability which was studied with a toy
model of scalar tachyons (Sect.~3.2) one may expect that the main result
of this study presenting the ground state of the tachyon field as a
superposition of zero-energy tachyons propagating isotropically will be
conserved for tachyon fields of any tensor dimensionality since the equation
(3.8), as well as the initial condition (3.6) leading to that particular 
solution, are expected to be valid for all individual tensor components of 
the fields. The only principal change which might appear due to replacement 
of a complex scalar field by an infinite-component one when constructing the 
tachyon vacuum wave function is a doubling of tachyon pairs in the coherent
mechanism of the tachyon vacuum creation required by the angular momentum
conservation (see e.g. Fig.~5b for explanation).  

For the same reason the production of a real (non-vacuum) tachyon of a high 
helicity state back-to-back with an antitachyon (generally speaking, with 
essentially non-zero opening angle) is suppressed in any reaction, unless the 
antitachyon is produced in parallel with the tachyon, with the helicities 
of both particles compensating each other. The overall angular momentum 
of such a pair can be low (even zero), but the pair mass squared in this 
case cannot exceed $-4 \mu^2$ restricting the production phase space.
Taken together, these properties can lead to a significant suppression of
the interactions of tachyons with ordinary particles, even if the tachyon 
coupling to them could be rather high, e.g. that of the electromagnetic 
interactions, $\alpha$. In particular, one can expect a strong suppression of 
high-energy Cherenkov radiation from charged tachyons due to the angular 
momentum conservation. Furthermore, an additional suppression of tachyon 
production amplitudes may turn out to be surprisingly high as a consequence of 
a naturally appearing conjecture about longitudinal non-locality of tachyons 
considered in the next subsection\footnote{The term ``longitudinal non-locality
of tachyons" is introduced to distinguish it from the ``spherical non-locality"
of scalar tachyons noticed by G. Feinberg in \cite{fein}.}. 

\subsubsection{Non-locality of tachyons}
While eigenvalues of the generator $M_{xy}$ are well-known quantum numbers
called helicities, the physical meaning of the generators $M_{xt}$ and $M_{yt}$
presents a novel feature. In the language of Lorentz transformation
in two transversal directions, they would leave invariant 
{\em a longitudinal tachyon size (i.e. the tachyon length)}, 
if we admit extended tachyons possessing such a property.
The same is true for rotations in the $xy$ plane. Thus this length will be 
an invariant of the $O(2,1)$ group, and we can indeed introduce a new quantum 
number, the tachyon intrinsic length. This is the way of how an idea of 
tachyons being explicitly non-local objects is emerging.

Let us call the tachyon intrinsic length by $l_0$. Since the $l_0$ cannot be
an independent invariant of the $O(2,1)$ group, it has to be related
to the Casimir invariant $Q$ of the $D_s^{\pm}$ branches or, more 
appropriately, to the value $s$, the minimal absolute tachyon helicity. 
Let us consider, for example, 
a possible expression for the ``visual appearance" of the tachyon length 
$l = s /(\chi E_t)$, where $E_t$ is a tachyon energy and the dimensionless
factor $\chi$ is introduced in order to account for our ignorance 
of tachyon characteristics. From a phenomenological consideration 
of the processes of tachyon-antitachyon pair production by high energy protons,
given in the next Section, one can conclude that the parameter $\chi$ can be 
within the range of $10^{-1} - 10^{-3}$, 
i.e. it may turn out to be not immensely low.

Thus the tachyon spatial extension has to be characterized by two parameters,
by $l_0$ as an intrinsic length, and by $\rho$ as a transversal size, 
as it should be for any axially-symmetric object. For the reasons mentioned
above (smallness of $\chi$) one has to assume that the tachyon length 
$l_0 >> r_\pi$, where $r_\pi$ is a typical hadronic radius, 1.4~fm,
while a vanishing size for the tachyon transverse dimension looks the most
natural\footnote
{If the condition $\rho << l_0$ is adopted
(including vanishing $\rho$), the size $\rho$ has no influence on the 
experimental characteristics of a tachyon; for example, it can be 
neglected in calculations of the tachyon Cherenkov radiation rate.}.
Then the resulting hierarchy of tachyon sizes displays tachyons as extended
stringlike objects with the string {\em extension, coupled to the tachyon 
helicity}, being directed strictly along the tachyon 
momentum\footnote{For the sake of completeness, we have to note that 
an axially-symmetric stringlike state with apparently superluminal group 
velocity can be obtained as a solution of a nonlinear Klein-Gordon equation 
\cite{terl}; however it is not related to UIR's of the Poincar\'{e} group.}.
Let us note however, that the tachyonic ``string" is, perhaps, a simpler object
than the open (as well as closed) strings of the standard string theory, 
in spite of the fact that both are non-local, one-dimensional objects. 
Since the tachyon extension is oriented strictly along its momentum, free 
motion of the tachyonic string can be described by a one-dimensional line 
in space (tending to a one-dimensional world line at $v \rightarrow c$), 
while free motions of the standard strings are described by two-dimensional 
world sheets.  

To conclude the presentation of tachyons as realization of the 
infinite-dimensional UIR's of the Poincar\'{e} group, let us summarize briefly 
their properties deduced from the consideration of its little group $O(2,1)$, 
corresponding to the spacelike particles, restricting ourselves to the
$D_s^\pm$ series. First of all, tachyons appear as extended (one-dimensional), 
axially-symmetric, stringlike objects. Their spins are directed along their 
momenta, to be more properly defined as helicities, always non-zero, as 
mentioned above. Ascribing to tachyons positive helicities, which may be 
either integer or half-odd-integer, the antitachyons will obtain negative 
ones. Tachyons can only be produced in pairs with antitachyons, conserving the 
tachyon number. A hint on the size of the tachyon longitudinal dimension 
(tachyon elongation) can be obtained from a restriction on the energy loss 
experienced by high energy ordinary particles due to spontaneous emission 
of tachyon-antitachyon pairs imposed by observations (see next section). 

\section{Several notes concerning assumed tachyon interactions 
with ordinary particles}
\setcounter{equation}{0}
\renewcommand{\theequation}{5.\arabic{equation}}
Unless a reasonable tachyon theory will be built one has to avoid the
consideration of possible {\em dynamic} effects which can be induced by 
tachyons to interaction amplitudes of ordinary particles giving rise to 
deviation from the Standard Model, 
for example, via appearance of tachyon loops in Feynman diagrams. 
Hopefully, these deviations are expected to be non-violating the
Lorentz invariance in the non-tachyon sectors of the theory, as noticed 
in Sect.~3.2. Moreover, they may be invoked for an explanation of some
existing discrepancies between theoretical predictions and experimental 
results for highly accurate calculations and measurements of radiative 
correction effects, e.g. those in the muon anomalous magnetic moment 
\cite{pdgmuamm}.

However, there are effects in which {\em kinematic} tachyon effects should 
play a dominant role, while the dynamics of tachyon interactions with ordinary 
particles can be represented by simple assumptions. In particular, 
owing to the gauge for the tachyon vacuum (2.25) one can speculate that a
theory of interacting tachyons promises to be rather similar, in principle, 
to that of ordinary particles, and tachyon interaction amplitudes can be 
constructed in a standard way applied to the construction of amplitudes of 
interactions between ordinary particles, of course, with the 
modifications necessary to account for the non-locality of tachyons. 
This conjecture will be used implicitly in all the considerations below
concerning several effects of this type.

\subsection{Spontaneous emission of a tachyon-antitachyon pair by a charged
particle}
An ordinary massive particle, being at rest in the preferred reference frame, 
cannot emit real tachyons since it is forbidden by the energy conservation
law (let us not forget that in the preferred reference frame 
no negative-energy tachyons or antitachyons can exist).
However this process, i.e. an anomalous decay of a particle to itself with the
emission of positive-energy tachyons, is kinematically allowed for particles 
moving with respect to the preferred reference frame \cite{kostya2,liebscher}. 
The only condition for a tachyon system to be emitted by such particles (the  
system must contain at least one tachyon-antitachyon pair since the production 
of a single tachyon pertaining to $D_s^\pm$ UIR's in any reaction with ordinary 
particles is forbidden) is the requirement for the spacelikeness of the system
4-momentum. Then a question may arise why the LHC works 
and why the high energy cosmic rays (HECR) exist? 

Rejecting a straightforward solution of this problem by appealing to an 
extremely weak coupling of tachyons to ordinary particles, the reasonable 
answer to this question can be obtained only assuming an essential
non-locality of the tachyon-antitachyon system, or more specifically, assuming 
a longitudinal extension of that system much bigger than the characteristic 
sizes of the ordinary particles. As has been shown above, this assumption can 
be easily accommodated with our consideration of tachyons as realizations of 
$D_s^\pm$ UIR's. Then the processes of the anomalous decays of particles with 
the emission of tachyons will be strongly suppressed due to a weak overlapping 
of the tachyon wave functions with those of the parent particles.
 
Let us consider, for example, the process of spontaneous emission of 
a tachyon-antitachyon pair by a proton possessing high energy in the
preferred reference frame, i.e. the reaction
\begin{equation}
   p_{in} \rightarrow p ~ t^+ t^-
\end{equation} 
and assuming tachyons to participate in standard electromagnetic interactions 
with ordinary particles, i.e. with an electromagnetic coupling constant 
$\alpha$ for such interactions. 

First we have to design a mechanism for such a reaction since it has no
analog in processes with ordinary particles\footnote{There is one exclusion
concerning the photon decay to an odd number of lower energy photons,
all the photons moving in the same direction (the reaction respects both, 
the Furry theorem and the energy-momentum conservation law, but has zero phase
space volume).}.   

Any asymptotic state (a proton in our case) is a wavepacket, i.e. a coherent
superposition of plane waves with some weight function. Interactions destroy 
the coherence. So, an inelastic particle collision can be considered as a 
creation of a highly incoherent fireball from which free particles emerge
after some formation time has passed, necessary for the coherence 
re-establishing. But what destroys the coherence of a proton, freely moving 
in vacuum, in the case of its anomalous decay to itself and to tachyons, 
i.e. in the reaction~(5.1)?

The answer is obvious: vacuum tachyons. In the proton rest frame the proton
``sees" a flux of vacuum tachyons and antitachyons possessing non-zero 
energies given, in particular, by formulae (3.12), (3.13) if they move along 
the direction of the preferred reference frame motion. 
They can interact with the proton at rest transferring
to it some kinematically allowed portion of their 4-momenta. In the preferred
reference frame this will be viewed as a spontaneous emission of tachyons by
the proton with the slowing of the latter. In this frame the process can be
described as an interaction of the moving proton with the vacuum tachyons
promoting them to become real ones. 

Within our hypothesis of the electromagnetic interactions of tachyons with 
ordinary particles the lowest order of the reaction (5.1) amplitude has to be 
of $\alpha^2$ due to a need of the local charge and momentum conservation 
and can be presented by a Feynman diagram displayed in Fig.~5a 
(as well as by an analogous diagram with the virtual photons
$q_1, q_2$ interchanged, and may be by some other diagrams of the same order
in $\alpha$, with photon lines between tachyons which would involve an 
additional non-local tachyon form-factor (tachyon wave function elongation), 
similar to those indicated in Fig.~5).

\vspace{1.0cm}
\includegraphics[height=7.5cm,width=24cm]{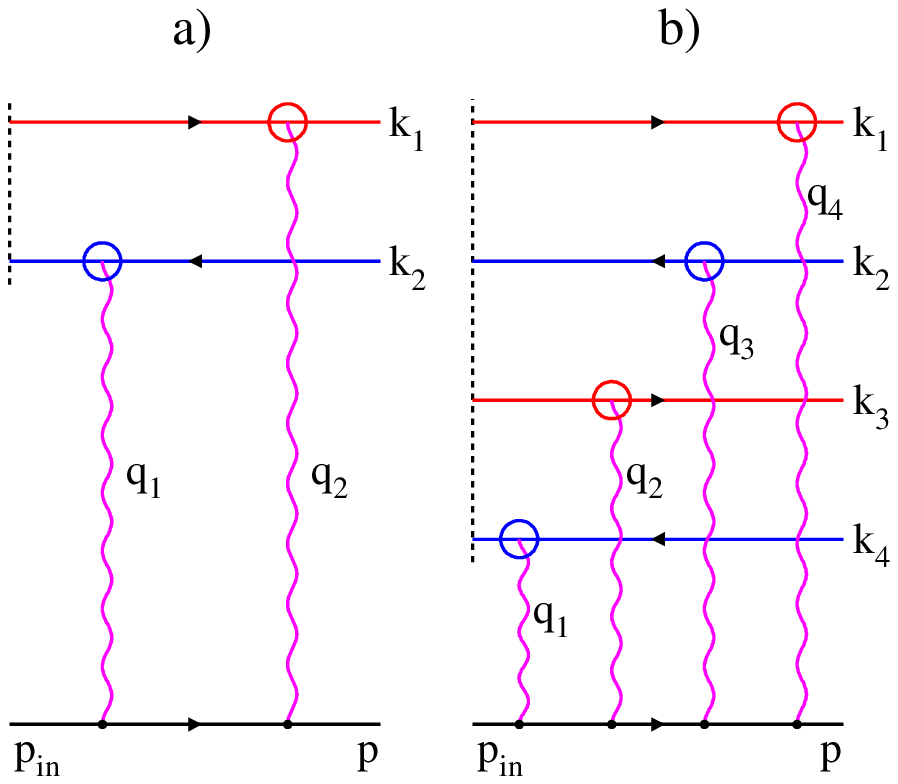}
Fig.~5. a) 2nd order Feynman diagram of tachyon pair production.            
b) 4th order Feynman diagram of two tachyon pair production. Open ellipses
at tachyon-photon vertices symbolize non-local tachyon form-factors (tachyon
wave function elongations). Vertical dashed lines symbolize the tachyon vacuum.
\vspace{5mm}

Moreover, the lowest order amplitude of interaction of a proton with vacuum 
tachyons may turn out to be of $\alpha^4$ as demonstrates the diagram in 
Fig.~5b, since the diagram 5a violates the law of the angular momentum 
conservation if the minimal tachyon helicity is greater than 1/2. But for the 
moment we shall restrict our consideration to the case of $\alpha^2$ 
interaction, i.e. of $\alpha^4$ in terms of probability. 

Each of the two tachyon-proton interaction sub-diagrams in Fig.~5a has to
contain a non-local tachyon form-factor, i.e. it will be proportional to a 
non-local term having (in the preferred reference frame) a form 
$\int d^3{\bf x}[\bar\psi({\bf x})\psi({\bf x})Q\bar\xi({\bf x})\xi({\bf x})]$, 
where $\psi({\bf x})$ and $\xi({\bf x})$ are the proton and tachyon wave 
functions and $Q$ is some matrix relating to the indices of the particle wave 
function components (hidden in the expression). The probability of the process 
will be given by a formula of a standard type:  
\begin{equation}
W  = \frac{(2\pi)^4}{2E_{in}}~\int |M|^2~d\tau,  
\end{equation}
where $E_{in}$ is the initial energy of the proton in the preferred reference
frame, M is a matrix element of reaction (5.1), and $d \tau$ is a phase 
space element. (Over)simplifying the situation, we present the $|M|^2$ by 
a constant term
\begin{equation}
|M|^2 = \alpha^4 \biggl(\frac{r_\pi}{l_0}\biggr)^4
\end{equation}
(most probably, the energy and angular dependences of the matrix element will 
decrease the probability).
The constant suppression factor 
\begin{equation}
\delta \sim \biggl(\frac{r_\pi}{l_0}\biggr)^4,
\end{equation}
comes from the weak overlap of the wave functions of the interacting particles 
in reaction (5.1): since the proton wave function differs from zero only in 
the region of dimension $r_\pi$ each (incoming and outgoing) tachyon 
brings a factor of $\sqrt{r_\pi /l_0}$ to the amplitude. 
Then, on the base of calculations presented in Appendix D, we obtain 
an approximate formula for energy loss by a high energy proton due to its
anomalous decay to itself and to a tachyon-antitachyon pair:
\begin{equation}  
\frac{dE}{dx}=\frac{1}{2(2\pi)^3}~\alpha^4~\biggl(\frac{r_\pi}{l_0}\biggr)^4 
~\frac{\mu^4}{E_{in}^2 } \ln \frac{E_{in}}{\mu}
\end{equation}  
(an ultra-relativistic case is assumed here, $E_{in} >> m_p, E_{in} >> \mu$,
and $\mu \geq m_p$; a little bit more accurate formula for $dE/dx$ is given 
in Appendix D). 

For our estimations we suggest $\mu = m_p$ and $r_\pi/l_0 = 10^{-3}$. With a 
calculated value of $dE/dx \approx 10^{-17}$~GeV/cm this results in about 
0.03~eV of the mean energy loss by a 7~TeV LHC proton per each turn in the LHC 
ring which is negligible as compared to other energy losses (e.g. due to 
the synchrotron radiation a 7~TeV LHC proton looses $\approx 6.7$~keV per turn 
\cite{synchr}). 

Neglecting a logarithmic dependence on $E_{in}$ in (5.5) one can estimate 
a HECR proton path until it looses a half of its initial energy:
\begin{equation}
   \Delta x_{\frac{1}{2}} \approx 0.3 E_{in}\bigg{/}\frac{dE}{dx}.
\end{equation}
Using (5.6) we have obtained an estimation for a proton path, 
of a proton of the primary HECR with the energy, say, $10^{15}$ eV,
equal to 30~Mpc, which corresponds to a typical scale of the propagation 
length of the highest energy cosmic rays.

One can interpret the processes of type (5.1) of the energy loss by 
particles moving with respect to the preferred reference frame due to a 
spontaneous emission of tachyons as a specific ``bremsstrahlung" experienced 
by charged particles moving through the tachyonic vacuum presenting some kind 
of a medium, a tachyonic ``ether" as it has been called in the Introduction. 
Having an accurate theory of tachyon interactions and the parameter $l_0$
evaluated with a smaller uncertainty, it would be interesting to 
calculate the influence of this ``bremsstrahlung" on peculiar velocities of 
galaxies and to compare the distribution of these velocities obtained with 
the nearest galaxies to that of the high redshift ones if the calculated 
difference in the distributions induced by this ``bremsstrahlung", in addition 
to the standard cosmological deceleration (2.1), would be measurable by modern 
astronomical tools.


\subsection{Production of tachyons in collisions of ordinary particles}
Unlike the anomalous $decays$ of the ordinary particles with tachyon
emission, the suppression of the tachyon production in processes of 
$collisions$ of ordinary particles is expected to be much weaker
since no vacuum tachyons are necessary anymore to participate in the
production of real tachyons. They can emerge from the fireball created by
colliding projectiles like the ordinary particles, so the lowest order
amplitude of the tachyon pair production via electromagnetic interactions
can be of $\alpha$, and the production cross section matrix element squared,
in the constant term approximation, will be given by
\begin{equation}
|M|^2 = \alpha^2 \biggl(\frac{r_\pi}{l_0}\biggr)^2,
\end{equation}
thus the suppression factor for the tachyon pair production
being of the order of
\begin{equation}
\delta \sim \biggl(\frac{r_\pi}{l_0}\biggr)^2.
\end{equation}

\subsection{Tachyons and atomic physics}
A big longitudinal tachyonic size can be invoked also when addressing the
questions coming from the atomic physics, at the opposite side of the 
microworld energy range. Why various atomic physics effects such as the fine
structure of the atomic spectra, Lamb shift, and others are not influenced
by tachyons having characteristic sizes $l_0$ 
of the order of $10^{-10}$ cm or so? 
The answer lies just in the atomic energy scale, which is concentrated mainly
within the range from 1 to 20~eV. 
If mass parameters of tachyons exceed 100~MeV, the tachyon longitudinal sizes 
at these energies will be stretched by $\sqrt{v^2 -1}$ coefficients exceeding 
$5\times 10^{6}$, and the suppression factors attenuating any 
Lorentz-non-invariant influence of vacuum tachyons on the atomic physics 
effects are expected to be at the level of $10^{-19}$ or less; even in the case
of the tachyonic masses of the order of the electron one (i.e. about 1~MeV) 
the Lorentz-non-invariant effects induced by tachyons may be 
suppressed by factors of the order of $10^{-11}$ or less.    

However when an accurate tachyon theory will be available and the parameter 
$l_0$ will be evaluated more accurately, it would be worth
to consider the possibility of an experiment aimed at looking for possible 
decoherence effects induced by the tachyonic ``ether" resembling the famous 
Michelson experiment, but using an atom wave interferometer instead of the 
optical one and assuming the velocity of the motion with respect to the ether
to be $10^{-3}~c$, instead of 30~km/s of the Earth orbital motion in the
original Michelson experiment. 

\section{Experimental status of the tachyon hypothesis}
Due to the strong theoretical objection to faster-than-light particles 
that they apparently violate causality, depicted in Sect.~2, 
few experiments have been made to search for them. From several experiments
carried out between the 1970 and 1987 with low energy hadronic beams 
\cite{baltay,danb, pvodd}, which used rather realistic assumptions about 
tachyon behaviour in particle detectors, one can conclude that tachyons 
do not participate in strong interactions, unless some specific mechanism 
exists, suppressing the cross sections (probabilities in \cite{baltay}) 
of their production in hadronic reactions by 3-4 orders of magnitude. 

Several experiments looking for charged tachyon production in
electromagnetic interactions using radioactive sources have also been made
\cite{alv1,alv2,pvmich}. However, restrictive upper limits on the tachyon
production that were concluded from these experiments were obtained under
strong assumptions about tachyon behaviour in the respective experimental
set-ups, and therefore, in our view, are questionable.

Currently the data collected by the Collaboration DELPHI at the CERN $e^+ e^-$ 
collider LEP are analysed with the aim to look for 
tachyons possessing electromagnetic interactions. The analysis 
is based on the tachyon properties deduced from the considerations presented
in this note. 

\section{Concluding remarks on a general approach to the tachyon problems}
Many serious problems of any Lorentz-invariant tachyon theory, mentioned in 
the Introduction, originate from the fact 
that the tachyon mass hyperboloid is one-sheeted, and therefore 
positive-energy tachyons on mass shell can be converted to the negative-energy 
tachyons by a proper Lorentz transformation. Unfortunately, ingenious 
suggestion of the principle reinterpretation \cite{bds} which replaces 
negative-energy tachyons by positive-energy antitachyons does not solve the 
problems. They can be solved only by a replacement of the second postulate of 
special relativity about equivalence of all inertial frames by a softer demand
to a theory to be expressed in a covariant form in any such frame, even though 
one of them is the preferred frame. Then the {\em Lorentz-covariant cut} 
of the tachyon mass hyperboloid (namely, the gauge (2.25)) separates 
$invariantly$ tachyons from antitachyons allowing {\em Lorentz-covariant 
calculations of the particle interaction probabilities} which are the main 
quantitative theoretical outcome in particle physics. 

One can ask why the principle of relativity works perfectly in the case of
ordinary particles, but has to be replaced by a softer requirement of  
the physical law covariance when dealing with tachyons. The reason for this
is a difference in the properties of the corresponding vacua. The vacuum of
ordinary particles is an essentially quantum-mechanical object and reveals 
itself locally in the world of elementary particles, in the sense that it 
``works" in the closest vicinity of a particle, whatever results of this 
``work" are: Unruh-Davies effect, Lamb shift or a need for the renormalisation 
of a theory. Even in the case of its influence as a global entity, as in the 
example of the electromagnetic vacuum in the Casimir effect, its properties are
insensitive to the global features of the macroworld (i.e. our universe). 
On the contrary, the tachyon vacuum, obtained under a demand of its stability 
and consisting of on-mass-shell, infinite speed tachyons, can be treated even 
classically, and appears to be very sensitive to those features which require 
to be accounted for in this case, leading to a loss of the relativity principle,
i.e. to the appearance of the preferred reference frame. 

This and other modifications of certain quantum aspects of the tachyon theory
(e.g. tachyon non-locality combined with the representation of the tachyon 
field operators by solutions of infinite-component wave equations) make it more 
difficult for a technical treatment as compared to the theory of ordinary 
particles. It is natural and perhaps inevitable: would tachyons be realizations
of the simplest, spin-zero UIR's of the Poincar\'{e} group, they would be  
easily ``reachable" experimentally though much more controversial from 
the theoretical point of view. 

Currently neither theory nor experiment demand the existence of the
faster-than-light particles, and only few experimental facts exist, lying 
outside of the Standard Model, which probably can be attributed to tachyons.
However for the reasons which look obvious (note for example a tremendous 
interest of the physical and public communities to the OPERA 
results \cite{opera} though from the very beginning
they looked to be incompatible with the tachyon hypothesis) 
the investigation of the possibility of particles existing beyond
the light barrier is a must of the scientific research. If they do not exist
we have to understand why. The former arguments against such particles can be
circumvented in a way shown above and, thus, turn out to be invalid. 
On the other hand, an observation of particles moving with superluminal 
velocities would mean a discovery of a new world of the nature constituents;
leaving aside a possible practical use of such a discovery, one can notice
that it would allow the establishment of a new role of the invariant speed 
of light as an universal, non-penetrable speed barrier between the two particle
worlds. 
  
E.C.G. Sudarshan stated in \cite{as} that no reason really exists for not
investigating the possible existence of faster-than-light particles 
experimentally, which was mainly ignored by experimenters. However,
after several decades have passed, this remains a true appeal.
  
\section{Summary}
The main ideas suggested in this note for a consistent tachyon theory 
can be listed as follows:
\begin{itemize}
\item [i)]
A postulate of the preferred reference frame is mandatory in any tachyon theory
in order to ensure the causality conservation.
\item [ii)] 
The tachyon vacuum seen in the preferred reference frame is a sea of 
zero-energy, on-mass-shell tachyons moving isotropically; this vacuum is stable.
There is a rotational asymmetry of the tachyon vacuum in non-preferred frames 
ensuring the causality conservation.
\item [iii)]
Scalar tachyons can be neutral only (generally speaking, with very weak
coupling to ordinary particles).
\item [iv)]
If tachyons are realizations of infinite-dimensional UIR's of the Poincar\'{e} 
group\footnote{or its extension.} (infinite-spin tachyons) they have to be 
produced in pairs with antitachyons.
\item [v)]
Infinite spin tachyons must be axially symmetric objects which logically leads 
to the conjecture about the existence of an intrinsic longitudinal size of 
the tachyons. Thus the tachyon theory has to be a non-local one. The extended 
tachyons can be electrically charged; in the case of ``normal" electromagnetic 
interactions of tachyons with ordinary particles estimations for the tachyon 
(longitudinal) sizes can be obtained requiring an agreement with observational
facts, to be at the level of $10^{-10}-10^{-12}$~cm.            
\end{itemize}
\subsection*{Acknowledgements}
\vskip 3 mm
The author is grateful to Profs. K.~G.~Boreskov, F.~S.~Dzheparov, 
A.~A.~Grigoryan, O.~V.~Kancheli and S.~M.~Sibiryakov for fruitful discussions, 
and to Dr.~B.~French for the critical reading of the manuscript.
\newpage
\section*{Appendix A. Tachyon kinematics}
\setcounter{equation}{0}
\renewcommand{\theequation}{A.\arabic{equation}}
Faster-than-light particles were postulated in \cite{bds} possessing the 
following properties. They cannot traverse the light barrier and be
brought to rest in any reference frame. Therefore their rest mass is imaginary,
$m = i\mu$, mass squared is negative, $m^2 = -\mu^2$, which determines their
four-momentum, $P = (E,{\bf p})$ to be spacelike. Thus
$E^2 - {\bf p}^2 = -\mu^2$. Defining the particle velocity by 
{\bf v} = {\bf p}/E the formulae for its energy and momentum become:
\begin{equation}
E = \frac{\mu}{\sqrt{v^2 -1}}
\end{equation} 
\begin{equation}
{\bf p} = \frac{\mu {\bf v}}{\sqrt{v^2 -1}}
\end{equation} 
Thus, the energy and momentum of the faster-than-light particle are always real.
As $v$ approaches 1 both the energy and momentum unlimitedly grow. 
Contrary, with the velocity increase they decrease, the energy 
approaching zero at $v$ approaching infinity, and the momentum tending
to the finite value $\mu$ (this is the tachyon state used when constructing 
the tachyon vacuum, see Sect.~3.2). The sign of the energy can be changed 
by a suitable Lorentz transformation,   
\begin{equation}
E^\prime = \frac{E-{\bf pu}} {\sqrt{1 - u^2}} =\frac{E(1-{\bf vu})} 
{\sqrt{1 - u^2}},
\end{equation}
if ${\bf v u} > 1$, where ${\bf u}$ is the relative velocity of two reference 
frames. Simultaneously the sign of the time component of an interval
connecting given points on the tachyon world line is changed. 
A coherent explanation of these changes was suggested in
\cite{bds}, denoted as {\em the principle of reinterpretation}. Accordingly to 
this principle, a faster-than-light particle of negative energy moving 
backward in time should be interpreted as an antiparticle of positive energy 
moving forward in time and in the opposite spatial direction. This 
reinterpretation is analogous to that proposed by Dirac, St\"uckelberg, 
Wheeler and Feynman for negative energy electrons going backward in time 
to be interpreted as positive energy positrons going forward in time
\cite{dirac,stuck,feyn}.
    
\section*{Appendix B. Unitary irreducible representations of the Poincar\'{e} 
group for spacelike momenta}
\setcounter{equation}{0}
\renewcommand{\theequation}{B.\arabic{equation}}
In Wigner's paper \cite{wigner2} several classes of UIR's of the
Poincar\'{e} group were considered, the UIR's corresponding to particles
with spacelike momenta, $P^2 < 0$, among them. To classify the UIR's Wigner
defined so called ``little group", selecting from all possible momentum
vectors a definite one denoted by $P^0$. Then he defines the little group
as the group of all Lorentz transformations $L$ which leave $P^0$ invariant,   
\begin{equation}
L P^0 = P^0.
\end{equation}
In the case of $P^2 < 0$, the $P^0$ is taken to be parallel to the $z$ axis
with the transformations $L$ which leave invariant the form $t^2 - x^2 - y^2$
(or, equivalently, the form $E^2 - p_x^2 - p_y^2 $).
The corresponding little group is a non-compact group of rotations in 
2+1~dimensions, which is called $O(2,1)$ \footnote{To display a distinction 
with the case of $P^2 > 0$ we note that in that case the vector $P^0$ is taken 
along the time axis and the resulting little group is a compact group $O(3$) 
whose UIR's, such as $d$-functions of the angular momentum theory, are well 
known.}.

The Lie algebra of $O(2,1)$ contains three independent elements, a compact 
generator $M_{xy}$, and two non-compact generators $M_{xt}$ and $M_{yt}$. 
$M_{xy}$ generates spatial rotations around the $z$ axis, i.e. in a plane 
perpendicular to the tachyon momentum (it is called the $helicity$
generator), while $M_{xt}$ and $M_{yt}$ are the generators of boosts in 
two independent directions in this plane.
The generators obey the following commutation relations:
\begin{equation}
[M_{xy},M_{xt}] = ~~i M_{yt},
\end{equation}
\begin{equation}
[M_{xy},M_{yt}] =  -i M_{xt},
\end{equation}
\begin{equation}
[M_{xt},M_{yt}] =  -i M_{xy}.
\end{equation}
Eigenvalues of $M_{xy}$ are either integers or half-odd integers with the
corresponding UIR's being either single or double-valued, respectively.
Within a given representation the eigenvalues differ from one another by 
integers.

The Casimir invariant of the little group (which is, up to a factor of 
$1/\mu^2$, the ``internal" Casimir invariant of the Poincar\'{e} group) 
is defined by
\begin{equation}
Q = M_{xt}^2 + M_{yt}^2 - M_{xy}^2.
\end{equation}

Solving commutation relations (B.2) - (B.4) the following classes of UIR's
were found, characterized by the value of the Casimir invariant $Q$ and
by the set of eigenvalues of the generator $M_{xy}$ (by {\em the helicity
spectrum}):
\begin{itemize}
 \item [a).] Continuous class, integral case: $h = 0, \pm  1, \pm 2,~..., 
ad ~inf.$, $0 < Q < \infty$.
 \item [b).] Continuous class, half-integral case: $h = \pm 1/2, \pm 3/2,
~..., ad ~inf.$, $1/4 < Q < \infty$.   
 \item [c).] Discrete class; it contains two branches: one, denoted by $D_s^+$, 
has positive helicities, $h_t = s, s+1, s+2,~..., ad ~inf.$, so $s$ is 
the lowest eigenvalue of $M_{xy}$, $s$ can assume the values 
$1/2, 1, 3/2, 2, ~..., ad ~inf.$, and another, denoted by $D_s^-$, with the 
same but negative helicities, so $-s$ is the highest eigenvalue of $M_{xy}$;
the representations of this branch are conjugate complex to the representations
of $D_s^+$. $Q = -s(s-1)$ for the both branches.    
 \item [d).] There is also a trivial solution of the commutation relations, 
$M_{xy} = M_{xt} = M_{yt} = 0$ resulting in trivial 
representations in which every group element is represented by the unit
operator. These representations correspond to spinless (scalar) tachyons. 
\end{itemize}

\section*{Appendix C. Quantization of a scalar tachyon field}
\setcounter{equation}{0}
\renewcommand{\theequation}{C.\arabic{equation}}
Consider, as a toy model, a free real scalar tachyon field with a field 
operator
\begin{equation}
\Phi(x) = \frac{1}{\sqrt{(2\pi)^3}} 
\int{d^4k~\Big{[}a(k)\exp{(-ikx)} + 
a^+(k)\exp{(ikx)}\Big{]}~\delta(k^2+\mu^2)~\Theta(kU)},
\end{equation}
where $k$ is a tachyon 4-momentum, $k = (\omega,{\bf k})$, ~$a(k), a^+(k)$ are
annihilation and creation operators, and $U$ is the 4-velocity of the preferred
 reference frame with respect to the observer. The presence of the delta 
function $\delta(k^2+\mu^2)$ in (C.1) is compulsory since we are considering 
free fields corresponding to mass shell particles; note, without it the theta 
function $\Theta(kU)$ is ill-defined. 

Application of $\delta(k^2+\mu^2)$ assumes, as usual, two roots of the 
equation $k^2+\mu^2=0$:
\begin{equation}
\omega = +\sqrt{{\bf k}^2 - \mu^2}
\end{equation}    
and
\begin{equation}
\omega = -\sqrt{{\bf k}^2 - \mu^2},
\end{equation}
both with $|{\bf k}| \geq \mu$, but in the preferred reference frame the root 
(C.3) is killed by the term $\Theta(kU)$. In a moving frame the yields of
(C.2), (C.3) are restricted by the same term to regions $\omega \geq {\bf ku}$. 
Therefore integrating (C.1) over $k^0$ gives, after expressing canonical
annihilation and creation operators $a_{{\bf k}}, a^+_{{\bf k}}$, annihilating 
or creating tachyonic states with 3-momentum ${\bf k}$, via $a(k), a^+(k)$
\begin{equation}
a_{{\bf k}} = a(k)~\Theta(kU)/\sqrt{2(kU)},
\end{equation}
\begin{equation}
a^+_{{\bf k}} = a^+(k)~\Theta(kU)/\sqrt{2(kU)},
\end{equation}
with the factors included to ensure a proper covariant normalisation of a 
single-tachyon wave function,
\begin{equation}
\Phi(t,{\bf x}) = \int_{|{\bf k}| > \mu,\omega > {\bf ku}}
{\frac{d^3{\bf k}}{2\omega}} \sqrt{\frac{2(\omega - 
{\bf ku})}{(2\pi)^3 \sqrt{1-u^2}}}
~\Big{[}a_{{\bf k}}\exp{(-i\omega t + i{\bf kx})} + 
a^+_{{\bf k}}\exp{(i\omega t - i{\bf kx})}\Big{]}.
\end{equation}
In the preferred reference frame
\begin{equation}
\Phi(t,{\bf x}) = \frac{1}{\sqrt{(2\pi)^3}} 
\int_{|{\bf k}| > \mu,\omega > 0}
{\frac{d^3{\bf k}}{\sqrt{2\omega}}~
\Big{[}a_{{\bf k}}\exp{(-i\omega t + i{\bf kx})} + 
a^+_{{\bf k}}\exp{(i\omega t - i{\bf kx})}\Big{]}}.
\end{equation}

Let us make here a remark about the minimal value of the energy of an 
individual tachyonic mode (in the preferred reference frame), 
$\omega_{min} \rightarrow 0$. The 
``zero-energy" state of a real tachyon has to be regarded as a state of very 
small but finite energy thus separating real tachyon states from the vacuum 
ones. This small (minimally detectable) energy is determined by the uncertainty 
principle, $\omega_{min} \Delta t \geq 1$, with $\Delta t = \Delta x/v$ 
where $\Delta x$ is the size of the experimental setup subdetector in which 
the tachyon energy can be measured, and $v$ is a tachyon velocity, i.e.  
$\Delta t=\Delta x~\omega_{min}/|{\bf k}|_{min} \approx\Delta x~\omega_{min}/\mu$ 
from which $\omega_{min} \geq \sqrt{\mu/\Delta x}$ (a similar estimation for
$\omega_{min}$ can be obtained from the uncertainty relation connecting 
$\Delta x$ with the tachyon momentum uncertainty at $|{\bf k}| \rightarrow \mu$)
\footnote{A typical value of $\Delta x$ of modern experimental installations 
can be estimated to be of order of $1 m$ (if tachyons participate in the 
electromagnetic interactions with the standard coupling $\alpha$). 
Then taking $\mu \approx 1$~GeV one obtains $\omega_{min} \geq 14$~eV.}. This
explains why equality signs are omitted in the integration limits in (C.6), 
(C.7) and in analogous limits below~\footnote{Another approach to the problem
of distinguishing real tachyons from the vacuum ones can be based on an 
assumption about quantitative distinction between them. A real tachyon should 
be represented by a wave packet of a final extension (length), while the vacuum
tachyons appear to be dispersed along their trajectories of an infinite length 
(indeed, over cosmological distances).}. 

Requiring the field (C.1) to obey the translational invariance the following
equation should hold:
\begin{equation}
 [P_\mu, \Phi(x)] = -i \partial_\mu \Phi(x), 
\end{equation}
where $P_\mu$ is an operator of a 4-momentum of the field. Its solution 
for  $P_\mu$ is:  
\begin{equation}
P_\mu=\frac{1}{2}\int{\frac{d^4k}{(2\pi)^3}~k_\mu~[a^+(k)a(k)+a(k)a^+(k)]~\delta(k^2+\mu^2)~\Theta(kU)}
\end{equation}
when choosing the bosonic commutation relations for $a, a^+$ operators:
\begin{equation}
[a(k), a(k^\prime)] = 0, ~~~[a^+(k), a^+(k^\prime)] = 0.
\end{equation}
\begin{equation}
[a(k), a^+(k^\prime)]~\delta(k^2 +\mu^2)~\delta(k^{\prime 2}+\mu^2)~
\Theta(kU)~\Theta(k^\prime U) = 
\delta^4 (k - k^\prime)~\delta(k^2 +\mu^2)~\Theta(kU).
\end{equation}
In particular, the field Hamiltonian is
\begin{equation}
H \equiv P^0 = \frac{1}{2}\int{\frac{d^4k}{(2\pi)^3}~k^0~[a^+(k)a(k)+a(k)a^+(k)]~\delta(k^2+\mu^2)~\Theta(kU)},
\end{equation}
which, after dropping as usually the infinite $c$-number related to zero-point
oscillations, results in 
\begin{equation}
H=\int_{|{\bf k}| > \mu,\omega>{\bf ku}}
{\frac{d^3{\bf k}}{(2\pi)^3}\frac{\omega-{\bf ku}}{\sqrt{1-u^2}}
~a^+_{{\bf k}} a_{{\bf k}}}.
\end{equation}
Thus the Hamiltonian is bounded from below
and is Hermitian. This is in an agreement with the statement formulated 
at the end of Sect.~4.2 that a tachyon theory with the gauge (2.25) for 
the tachyon vacuum can be made unitary, just due to a corresponding property 
of the time evolution operator $\exp{(-iHt)}$.  

In the preferred reference frame
\begin{equation}
H = \int_{|{\bf k}| > \mu,\omega>0}
{\frac{d^3{\bf k}}{(2\pi)^3}~\omega ~a^+_{{\bf k}} a_{{\bf k}}}
\end{equation}
having non-negative eigenvalues and resembling, in accordance with the 
correspondence principle formulated at the
end of Sect.~2, the standard expression for ordinary scalar particles. 

Let us express the Hamiltonian in terms of the field $\Phi$. First, using (C.6)
we write $a({\bf k}), a^+({\bf k})$ in terms of $\Phi, \dot\Phi$. On the
spacelike hypersurface $t = t_0$ they are:
\begin{equation}
a_{{\bf k}} = \frac{\omega (1 -u ^2)^{1/4}}
{\sqrt{2(\omega - {\bf ku})}}~\exp{(i\omega t_0)}
\int{d^3{\bf x}~\exp{(-i{\bf kx})}~\Biggl{[}\Phi(t_0,{\bf x}) + 
\frac{i\dot\Phi(t_0,{\bf x})}{\omega}\Biggr{]} }, 
\end{equation}
\begin{equation}
a^+_{{\bf k}} = \frac{\omega (1 - u^2)^{1/4}}
{\sqrt{2(\omega - {\bf ku})}}~\exp{(-i\omega t_0)}
\int{d^3{\bf x}~\exp{(i{\bf kx})}~\Biggl{[}\Phi(t_0,{\bf x}) - 
\frac{i\dot\Phi(t_0,{\bf x})}{\omega}\Biggr{]} }.
\end{equation}
Inserting (C.15), (C.16) into (C.13) we get 
\begin{equation}  
H = \int_{|{\bf k}| > \mu,\omega>{\bf ku}}
{\frac{d^3{\bf k}~\omega^2}{2(2\pi)^3}
\int{d^3{\bf x} d^3{\bf y}\Biggl{[}\Phi(t_0,{\bf x}) - 
\frac{i\dot\Phi(t_0,{\bf x})}{\omega}\Biggr{]} 
\Biggl{[}\Phi(t_0,{\bf y}) + 
\frac{i\dot\Phi(t_0,{\bf y})}{\omega}\Biggr{]}}\exp{[-i{\bf k(x-y)}]} }
\end{equation}
Noting that 
$\int_{|{\bf k}| > \mu,\omega>{\bf ku}}{d^3{\bf k}~\exp{[-i{\bf k(x-y)}]}} =
(2\pi)^3\bar\delta^3({\bf x-y})$ and\\
$\int_{|{\bf k}| > \mu,\omega>{\bf ku}}
{d^3{\bf k}~\omega^2 \exp{[-i{\bf k(x-y)}]}} =
(2\pi)^3 [-(\partial/\partial x_i)^2 - \mu^2]~\bar\delta^3({\bf x-y})$,
where $\bar\delta^3 ({\bf x-y})$ is a ``truncated" delta function which acts
like the standard delta function with respect to those functions whose Fourier
transforms vanish at $|{\bf k}| < \mu, ~\omega < ({\bf ku})$, we obtain finally
the Hamiltonian 
\begin{equation}
H = \frac{1}{2} \int d^3 {\bf x} \Big{[}\dot\Phi^2(x) +
\nabla\Phi(x) \nabla\Phi(x)  - \mu^2 \Phi^2(x) \Big{]} 
\end{equation}
which is local, as distinct to that of the Feinberg's model, and corresponds 
to a local Lagrangian
\begin{equation}
L = \frac{1}{2} \int d^3 {\bf x} \Big{[}\dot\Phi^2(x) -
\nabla\Phi(x) \nabla\Phi(x)  + \mu^2 \Phi^2(x) \Big{]}
\end{equation}
allowing a standard Lagrangian formalism of the field description to be used.

Let us compare the Lagrangian (C.19) with that in the expression (3.14)
(ignoring for the moment the distinction between real and complex fields in 
these expressions). They differ by a Lorentz-non-invariant term presented in 
the latter. This term has been introduced to (3.14) to describe the deviation 
of the theory in the tachyon sector from the Lorentz invariance motivated by
the requirement of the causality conservation, while the 
starting point of the approach developed in Sect.~3.2 was a Lorentz-invariant 
Lagrangian~(3.1). Contrary, in this Appendix we started with an explicitly 
Lorentz-non-invariant tachyon field operator and arrived at an apparently 
Lorentz-invariant Lagrangian. Why did such metamorphoses happen?  

The reason is the fact that, in analogy with the argumentation for introducing 
the Lorentz-non-invariant term to (3.14), in the case of the real scalar field 
under consideration we can add to the Lagrangian a similar term 
$\lambda U^\mu \partial_\mu\Phi(x)$. Because this additional term, written 
down as $\lambda \partial_\mu F^\mu(x)$, where $F^\mu(x) \equiv U^\mu \Phi(x)$,
is proportional to the total divergence of the 4-vector $F^\mu(x)$, the both
Lagrangians, with and without the additional term, are physically equivalent
since the term with $\partial_\mu F^\mu(x)$ does not contribute to physical
quantities excepting those related to the tachyon vacuum.  

Furthermore, this additional term, as well as the Lorentz-non-invariant 
term in (3.14), does not change the tachyon equation of motion (3.8). 
Therefore within our approach the Lorentz-invariance can be defined as 
spontaneously broken and its violation appears to be restricted to 
the asymptotic-tachyon-states sector only, even in the case of presumed 
tachyon interactions with ordinary particles. Considering a tachyon 
propagator as an inverse of the wave equation (3.8) in momentum space, 
we can write down, for example, the Feynman propagator as
\begin{equation}
\tilde G_F(k) = \frac{i}{k^2 + \mu^2 + i\epsilon}  
\end{equation}
to be used in Feynman diagrams describing tachyon interactions, of course, 
only within our toy model of scalar tachyons. In the particle configuration 
space
\begin{equation}
G_F(x-y) = \int_{|\bf{k}|\geq \mu} {\frac{d^4 k} {(2\pi)^4}~ 
\frac{i~\exp{[-ik(x-y)]}} {k^2+\mu^2+i\epsilon}}. 
\end{equation}
The standard $i\epsilon$ prescription makes the tachyon Feynman propagator
Lorentz invariant likewise the Feynman propagators of ordinary particles.
Contrary, the retarded and advanced tachyon Green functions of classical 
tachyon field theory are not invariant and can be transformed one to another 
in a specific kinematic domain by a suitable Lorentz transformation,
which agrees well with the Bilaniuk-Desphande-Sudarshan reinterpretation
principle \cite{bds}.

An important item are commutation relations for the tachyon field operators 
$\Phi, \dot\Phi$. The commutation relations (C.10), (C.11) for the operators 
$a, a^+$ lead to the commutation relation for the tachyon fields: 
\begin{equation}
[\Phi(x),\Phi(y)] = \int{\frac{d^4k}{(2\pi)^3}~\Big{\{}}{\exp[-ik(x-y)] - 
\exp[(ik(x-y)]\Big{\}}~\delta(k^2+\mu^2)~\Theta(kU)}
\end{equation}
which is not automatically zero at $(x-y)^2 < 0$ as distinct to the field 
commutators of ordinary particles. Consider it in the preferred reference frame:
\begin{equation}
[\Phi(x),\Phi(y)] = \frac{1}{(2\pi)^3}
\int_{|{\bf k}| > \mu,\omega > 0}{\frac{d^3{\bf k}}{2\omega}~
\Big{\{}\exp{[-i \omega \Delta t + i{\bf k(x-y)]}}
-\exp{[i\omega \Delta t - i{\bf k(x-y)]}}\Big{\}}},
\end{equation}
where $\Delta t = x^0 - y^0$. If $\Delta t \neq 0$ the commutator does not 
vanish; moreover, with this condition in the preferred reference frame it does 
not vanish in any other frame since the expression (C.22) is Lorentz-covariant. 

On the other hand, if $\Delta t = 0$ in the preferred reference frame
the commutator (C.23) vanishes:
\begin{equation}
[\Phi(x),\Phi(y)]_{x^0=y^0}=\frac{1}{(2\pi)^3}
\int_{|{\bf k}| \geq \mu,\omega \geq 0}
{\frac{d^3{\bf k}}{2\omega}~
\Big{\{}\exp{i{\bf k(x-y)}} -\exp{-i{\bf k(x-y)}}\Big{\}}} = 0
\end{equation}
which is a nice feature corresponding to the impossibility of superluminal 
communications with the infinite speed of the signal 
(${\bf v} = ({\bf x - y})/\Delta t = \infty$ in this case), 
i.e. via exchange by zero-energy (vacuum\footnote{See footnote 23.}) 
tachyons which is intuitively obvious. 
Note that both Feinberg~\cite{fein} and Arons-Sudarshan~\cite{as} 
models do not possess this property: in their models the field equal-time 
(anti)commutators do not vanish:
\begin{equation}
\{\Phi(x),\Phi(y)\}_{x^0=y^0} = \frac{1}{(2\pi)^3} \int
{\frac{d^3{\bf k}}{\omega}~\exp i{\bf k(x-y)}},
\end{equation}  
see formula (4.14) (for a real field) in \cite{fein}, and 
\begin{equation}
[\Phi(x),\Phi^+(x^\prime)]_{x^0=x^{0\prime}} = \frac{1}{(2\pi)^3} \int
{\frac{d^3{\bf k}}{\omega}~\exp i{\bf k(x-x^\prime)}},
\end{equation}
formula (3.8) (for a complex field) in \cite{as}. 

A similar complaint can be addressed to Feinberg's 
$\{\dot\Phi(x),\dot\Phi(y)\}_{x^0=y^0}$, non-equal to zero, see expression 
(4.17) in \cite{fein}. In our model 
\begin{equation}
[\dot\Phi(x),\dot\Phi(y)]_{x^0=y^0} = 0,
\end{equation}
while the equal-time commutator of the field $\Phi$ with its canonical 
conjugate $\dot\Phi$ does not vanish corresponding to the analogous commutator 
for the ordinary particle fields:
\begin{equation}
[\Phi(x),\dot\Phi(y)]_{x^0=y^0} = i~\bar\delta^3 {\bf(x-y}). 
\end{equation}

Such are the main features of a free scalar real field of faster-than-light 
particles in our model (one can note that it shares several common properties 
with the model \cite{ds}, c.f. commutator relations (C.24), (C.28) here 
and those (1.10) in \cite{ds}). 

The generalization of the method to a complex scalar field is 
straightforward:
\begin{equation}
\Phi(x) = \frac{1}{\sqrt{(2\pi)^3}} 
\int{d^4k~\Big{[}a(k)\exp{(-ikx)} +  
b^+(k)\exp{(ikx)}\Big{]}~\delta(k^2+\mu^2)~\Theta(kU)}, 
\end{equation}
\begin{equation} 
\Phi^+(x) = \frac{1}{\sqrt{(2\pi)^3}} 
\int{d^4k~\Big{[}a^+(k)\exp{(ikx)} + 
b(k)\exp{(-ikx)}\Big{]}~\delta(k^2+\mu^2)~\Theta(kU)} 
\end{equation}
with the commutation relations for $a, a^+$ and $b, b^+$ being similar to 
(C.10), C(11).

The field commutator
\begin{equation}
[\Phi(x),\Phi^+(y)] = \int{\frac{d^4k}{(2\pi)^3}~\Big{\{}}{\exp[-ik(x-y)] - 
\exp[(ik(x-y)]\Big{\}}~\delta(k^2+\mu^2)~\Theta(kU)}
\end{equation}
reproduces (C.22) with all its properties; in particular, the nonvanishing
commutator (C.31) means that the propagation of a real tachyon (across a 
spacelike interval) is distinguishable from the propagation of an antitachyon 
in the opposite spatial direction.

Similarly, the Hamiltonian and 
Lagrangian of the field are given, up to a factor of 1/2, by expressions
(C.18) and (C.19), with the bilinear forms in $\Phi$ being replaced by those
in $\Phi, \Phi^+$. There exists, as expected, a charge-current 4-vector  
\begin{equation}
j^\mu = i (\Phi^+ \partial^\mu\Phi - \partial^\mu\Phi^+ \Phi)
\end{equation}
which satisfies the usual continuity equation. The total charge of the field is
\begin{equation}
Q \equiv \int d^3 {\bf x}~j^0({\bf x}) = \int_{|{\bf k}| > \mu,\omega>{\bf ku}}
{\frac{d^3{\bf k}}{(2\pi)^3}\frac{\omega-{\bf ku}}{\omega \sqrt{1-u^2}}~
\Big{[}a^+_{{\bf k}} a_{{\bf k}} - b_{{\bf k}} b^+_{{\bf k}}\Big{]}}
\end{equation}
which is an explicitly Lorentz-covariant value. Defining the numbers of 
tachyons and antitachyons by
\begin{equation}
N_t = \int_{|{\bf k}| > \mu,\omega>{\bf ku}}
{\frac{d^3{\bf k}}{(2\pi)^3}~\frac{\omega-{\bf ku}}{\omega \sqrt{1-u^2}}~
a^+_{{\bf k}} a_{{\bf k}}},
\end{equation}
\begin{equation}
N_{\bar t} = \int_{|{\bf k}| > \mu,\omega>{\bf ku}}
{\frac{d^3{\bf k}}{(2\pi)^3}~\frac{\omega-{\bf ku}}{\omega \sqrt{1-u^2}}~
b^+_{{\bf k}} b_{{\bf k}}}
\end{equation}
we obtain, up to a standard infinite additive constant,
\begin{equation}
Q = N_t - N_{\bar t}~.
\end{equation}

\section*{Appendix D. Estimation of energy loss by high energy protons
via emission of tachyon-antitachyon pairs} 
\setcounter{equation}{0}
\renewcommand{\theequation}{D.\arabic{equation}}
Representing the matrix element squared of the reaction (5.1) probability
by a constant term $\alpha^4\bigl(\frac{    r_\pi}{    l_0}\bigr)^4$
the proton energy loss can be approximated by
\begin{equation}
\frac{dE}{dx}=\frac{(2\pi)^4}{2E_{in}}~
\alpha^4\biggl(\frac{    r_\pi}{    l_0}\biggr)^4 \int(E_{in}-E)
\frac{d^3{\bf p}}{2E(2\pi)^3}~
\frac{d^3{\bf k_1}}{2\omega_1(2\pi)^3}~\frac{d^3{\bf k_2}}{2\omega_2(2\pi)^3}~
\delta^4 (p_{in} - p - k_1 - k_2) 
\end{equation}
where 4-momenta of the initial and final protons are 
$p_{in} = (E_{in},{\bf p}_{in}),~p = (E,{\bf p})$, and 4-momenta of tachyons 
are $k_1 = (\omega_1, {\bf k}_1),~k_2 = (\omega_2, {\bf k}_2)$. 
After some algebra  
\begin{equation}
\frac{dE}{dx}=\frac{1}{2(2\pi)^3}~\alpha^4~\biggl(\frac{r_\pi}{l_0}\biggr)^4 
~\frac{\mu^4}{E_{in}|{\bf p}_{in}| } 
\ln \frac{E_{in}-E_{min} + \sqrt{(E_{in}-E_{min})^2 + 4\mu^2}}{2\mu},
\end{equation}
where 
\begin{equation}
E_{min} = E_{in}~\Biggl{\{}1  - \frac{2\mu}{m_p}~\Biggl[\sqrt{
\biggl(1+\frac{\mu^2}{m^2_p}\biggr)\biggl(1-\frac{m^2_p}{E^2_{in}}\biggr)}
-\frac{\mu}{m_p}~\Biggr]\Biggr{\}}
\end{equation}
with the threshold condition for $E_{in}$:
\begin{equation}
E_{in} \geq \sqrt{m^2_p + \mu^2}
\end{equation}


\end{document}